\documentstyle[12pt,jeep]{article}
\numberbysection
\date{}
\begin{document}
\title{Time Dependent Resonance Theory}
\author{
A. Soffer \thanks{Department of Mathematics, Rutgers University, New
Brunswick, NJ} \hspace{.05 in}
 and
 M.I. Weinstein \thanks{Department of Mathematics, 
University of Michigan, Ann Arbor, MI 
 }}
\baselineskip=18pt
\maketitle
\newcommand{\ra}{\rightarrow}
\newcommand{\da}{\downarrow}
\newcommand{\wra}{\rightharpoonup}
\newcommand{\rf}{\widehat}
\newcommand{\nit}{\noindent}
\newcommand{\nn}{\nonumber}
\newcommand{\be}{\begin{equation}}
\newcommand{\ee}{\end{equation}}
\newcommand{\ba}{\begin{eqnarray}}
\newcommand{\ea}{\end{eqnarray}}
\newcommand{\bom}{\mbox{$\Omega$}}
\newcommand{\Fee}{\mbox{$\Phi$}}
\newcommand{\del}{\mbox{$\nabla$}}
\newcommand{\delx}{\mbox{$\nabla_{x}$}}
\newcommand{\dely}{\mbox{$\nabla_{y}$}}
\newcommand{\ar}{\mbox{$\alpha$}}
\newcommand{\fee}{\mbox{$\varphi$}}
\newcommand{\dta}{\mbox{$\delta$}}
\newcommand{\lam}{\mbox{$\lambda$}}
\newcommand{\Lam}{\mbox{$\Lambda$}}
\newcommand{\eps}{\mbox{$\epsilon$}}
\newcommand{\gam}{\mbox{$\gamma$}}
\newcommand{\Gam}{\mbox{$\Gamma$}}
\newcommand{\al}{\mbox{$\alpha$}}
\newcommand{\bt}{\mbox{$\beta$}}
\newcommand{\va}{\mbox{$\bf a $}}
\newcommand{\db}{\mbox{$\parallel$}}
\newcommand{\ov}{\overline}
\newcommand{\ud}{\underline}
\newcommand{\nW}{\left| W\right|_{\bf a}}
\newcommand{\D}{\partial}
\newcommand{\px}{\mbox{$\partial_{xx}^{2}$}}
\newtheorem{theo}{Theorem}[section]
\newtheorem{defin}{Definition}[section]
\newtheorem{prop}{Proposition}[section]
\newtheorem{lem}{Lemma}[section]
\newtheorem{cor}{Corollary}[section]
\newtheorem{rmk}{Remark}[section]
\renewcommand{\theequation}{\arabic{section}.\arabic{equation}}
\def\R{{\rm \rlap{I}\,\bf R}}
\def\C{{\bf C} \llap{
 \vrule height 6.6pt width 0.6pt depth 0pt \hskip 4.7pt}}
\def\WB{W\tilde g_\Delta(H)}
\def\gI{g_I(H_0)}
\def\tg{\tilde g}
\def\tA{\tilde A}
\def\tB{\tilde B}
\newcommand{\nm}[1]{\vert\vert {#1} \vert\vert}
\def\la{\langle}
\def\ra{\rangle}
\def\jxs{\la x\ra^{\sigma}}
\def\jxms{\la x\ra^{-\sigma}}
\def\jt{\la t\ra}
\def\j#1{\la #1 \ra}
\def\ld{[\phi_d]_{LD}}
\def\eW{\eta_W}
\def\Xnm#1{\nm{#1}_X}
\begin{abstract}
An important class of resonance problems involves 
 the study of perturbations of
systems having embedded eigenvalues in their continuous spectrum. 
Problems with this mathematical structure arise in the study of many
physical systems, {\it e.g.} the coupling of an atom or molecule to a
photon-radiation field, and Auger states of the helium atom, as well as
in spectral geometry and number theory. 
We present a dynamic (time-dependent) theory of such {\it quantum resonances}.
The key hypotheses are (i) a resonance condition which holds generically
(non-vanishing of the {\it Fermi golden rule})
 and (ii) 
local decay estimates for the unperturbed dynamics with initial data
 consisting of continuum modes associated with an interval containing the
embedded eigenvalue of the 
 unperturbed Hamiltonian. No assumption of
dilation analyticity of the potential is made. 
Our method explicitly demonstrates the  
  flow of energy from the
resonant discrete mode to continuum modes due to their coupling.
The approach is also applicable to nonautonomous linear problems and
to nonlinear problems. 
We derive the time behavior of the resonant states for 
intermediate and long times.
  Examples and applications are
presented. Among them is a proof of the instability of an embedded
eigenvalue at a threshold energy under suitable hypotheses.
\end{abstract}
\thispagestyle{empty}
\vfil\eject
\bigskip
\vfil\eject

\section{Introduction}

The theory of resonances has its origins in attempts to explain the
existence of metastable states in physical systems. These are states
which are localized or coherent for some long time period, called
the {\it lifetime}, and then disintegrate. Examples abound and include
unstable atoms and particles.

The mathematical analysis of resonance phenomena naturally leads to the
study of perturbations of self-adjoint operators which have embedded
eigenvalues in their continuous spectra. An example of this is in the
quantum theory of the helium atom, in which there are long-lived Auger states
 \cite{kn:RS4}.
 The mathematical study of this problem proceeds by viewing 
as the unperturbed  self-adjoint operator, the Hamiltonian governing two
decoupled electron-proton systems. This system has many embedded
eigenvalues. The perturbed Hamiltonian is that which includes 
 the effect of electron-electron repulsion. In examples 3 and 4 of
section 6, we discuss a class of problems with this structure. Another
physical problem in which resonances play an important role is in the
setting of an atom coupled to the photon-radiation field 
\cite{kn:BFS}, \cite{kn:JP1}, \cite{kn:JP2}, \cite{kn:King92},
\cite{kn:King95}; see also example 7 of section 6.
   Although initially inspired by
the study of quantum phenomena, questions involving embedded eigenvalues
  have been seen to arise,
 quite naturally in spectral geometry and number theory \cite{kn:PhSa}.
The systematic mathematical study of the effects of perturbations on
embedded eigenvalues was initiated by Friedrichs \cite{kn:Fr}.

The method of analyzing the resonance problem we develop here is related
to our work on the large time behavior of nonlinear Schr\"odinger and
nonlinear wave equations {\cite{kn:SW2bs}, \cite{kn:SWrdamp}, \cite{kn:SW1bs}. 
\footnote{ Some of the results of this paper were presented  
 in  the proceedings article 
\cite{kn:SWres0} and in the preprint \cite{kn:SWres-dook}}
 In these problems, certain states of the
system decay slowly as a result of resonant interactions generated by
nonlinearity in the equations of motion. The methods 
required  are    
necessarily time-dependent as the equations are nonlinear and
nonintegrable. They are based on a direct approach to the study of
energy transfer from discrete to continuum modes. 
\footnote{
Related to this is the observation that
many nonlinear
phenomena can be regarded as (generic) instabilities of embedded eigenvalues for
 suitable {\it linear} operators. This point of view
 is taken by I.M. Sigal in \cite{kn:S1},\cite{kn:S2}, who studies the  
non-existence of
bifurcating time-periodic and
spatially localized solutions of certain nonlinear wave and
Schr\"odinger equations. 
 The problem of absence
of small amplitude {\it breathers}  
 for Hamiltonian perturbations of the
Sine-Gordon equation (see, for example, \cite{kn:SK} \& \cite{kn:BMW})
 can also be viewed in this context \cite{kn:S2}. Other nonlinear wave
 phenomena, in which resonances have been shown to play a role, are
 studied in \cite{kn:PW}, \cite{kn:CH}.}

We consider the following general problem. Suppose $H_0$ is a
self-adjoint operator in a Hilbert space ${\cal H}=L^2(\R^n)$, such that
$H_0$ has 
 a simple eigenvalue, $\lambda_0$, which is embedded in its continuous
spectrum, with associated eigenfunction, $\psi_0$:
$$ H_0\psi_0 =\lambda_0\psi_0, \qquad \vert\vert\psi_0\vert\vert_2=1.$$
We now consider the time-dependent Schr\"odinger equation, for the
perturbed self-adjoint Hamiltonian, $H=H_0 + W$,
\be i\partial_t\phi = H\phi \label{eq:tdeps} \ee
where $W$ is a perturbation which is small in a sense to
be specified. The choice of decomposition of $H$ into an unpertured part,
$H_0$, and a perturbation, $W$, depends on the problem at hand; see, for
example, \cite{kn:Demuth}.
\medskip

\noindent {\bf Problem:} {\it Suppose we specify initial data, $\phi_0$ for
(\ref{eq:tdeps})
which are spectrally localized (relative to $H$) in a small interval
$\Delta$ about
$\lambda_0$. Describe the
time-dynamics for $t\in (-\infty,\infty)$.}
\medskip

We shall prove that under quite general assumptions on $H_0$ and $W$
that for small perturbations $W$,

\noindent (i)  $H$ has absolutely continuous spectrum in an interval
about
$\lambda_0$,

\noindent (ii) the solution with such data decays algebraically
  as $t\to\pm\infty$.
For the special case of initial conditions given by 
 $\psi_0$,
the solution is characterized by transient exponential decay. 
 The exponential rate, $\Gamma$,
   (reciprocal of the lifetime) can be calculated.

On the more technical side, we have imposed fairly relaxed hypotheses on
the regularity of the perturbation, $W$; in particular we do not require
any condition on its commutators. This may be useful in problems like
the radiation problem and problems where Dirichlet decoupling is used.

The decay of solutions due to resonant coupling to the continuum
is revealed by decomposing the solution of (\ref{eq:tdeps}), with data
spectrally localized (relative to $H$) near $\lambda_0$, in terms of
the natural basis of the unperturbed problem:
\be
\phi(t)\ =\ a(t)\ \psi_0\ +\ {\tilde \phi}(t),\quad
\left(\psi_0,{\tilde\phi}(\cdot,t)\right)=0.\nn
\ee
After isolating the key resonant contributions, the system of equations
governing $a(t)$ and ${\tilde\phi}$ is seen to have the form:
\ba
ia'\ &=&\ (\Lambda\ -\ i\Gamma)a\ +\ {\cal C}_1(a,{\tilde\phi}) 
\nn\\
i\D_t{\tilde\phi}\ &=&\ H_0{\tilde\phi}\ +\ 
 {\cal C}_2(a,{\tilde\phi}),
\label{eq:model}
\ea
where the ${\cal C}_j,\ j=1,2$ denote terms which couple the dynamics of $a$ and
${\tilde\phi}$, and ${\cal C}_2$ lies in the continuous spectral part of
$H_0$. If these coupling terms are neglected, then it is clear that 
 $a(t)$ is driven to
zero provided $\Gamma>0$. The quantity, $\Gamma$, is displayed in
(\ref{eq:fgr}) and is always nonnegative. 
Its explicit formula, (\ref{eq:fgr}),  is often referred to as the {\it
Fermi golden rule}.
Generically, $\Gamma$ is strictly positive. 
 The exponential behavior suggested by these
heuristics is, in general, only a transient; 
in general, $e^{-iH_0t}$ has dispersive wave solutions, and coupling to
these waves leads to (weaker) 
algebraic decay as $t\to\pm\infty$. At this stage, we wish to point out
that although presented in the setting of a Schr\"odinger type operator,
acting in $L^2(\R^n)$, our results and the approach we develop below 
 can be carried out in the setting of a general Hilbert space, ${\cal
 H}$, with appropriate modifications made in the hypotheses. These
 modifications are discussed in the remark following our main theorem,
 Theorem 2.1. Their implementation is discussed in several of the
 examples presented in section 6.

Historically, motivated by experimental observations,
the primary focus of mathematical analyses of the resonance problem 
 has been on establishing exponential decay at intermediate
times. However, viewed as an infinite dimensional
Hamiltonian system,
  the asymptotic ($t\to\pm\infty$) behavior of solutions is
a fundamental question. Our methods address this question  and are
adaptable to nonautonomous linear, and nonlinear problems
\cite{kn:SW2bs},
\cite{kn:SWrdamp}.

The time decay of such solutions implies that the spectrum
of the perturbed Hamiltonian, in a neighborhood of $\lambda_0$,
  is absolutely continuous. 
 This implies the instability of the embedded eigenvalue. More
precisely, under perturbation the embedded eigenvalue moves off the real
axis and becomes a pole ("resonance pole" or "resonance energy") 
 of the resolvent analytically continued across
the continuous spectrum onto a second Riemann sheet \cite{kn:H}. 
  We will also
show that in a neighborhood of such embedded eigenvalues, there are no 
 {\it new} 
embedded eigenvalues which appear, and give an  estimate on the size
of this neighborhood.  Most importantly, 
we find the time behavior of solutions of the associated 
Schr\"odinger type evolution equation for short, intermediate and long time
scales. The {\it lifetime} of the resonant state naturally emerges from
our analysis. These results are stated precisely in Theorem 2.1.

Many different approaches to
the resonance problem in quantum mechanics 
 have been developed over the last 70 years and the various
 characterizations of resonance energies are expected to be equivalent;
 see \cite{kn:HeSj}.
The first (formal) approach to the resonance problem, due to Weisskopf \& Wigner \cite{kn:WW},
was introduced in their study of the phenomena of spontaneous emission
 and the instability of excited states; see also \cite{kn:Landau}.    
 Their approach  plays a central role in today's physics literature; 
 see for example
\cite{kn:AE}, \cite{kn:LL}.  It
is time-dependent and our approach is close in spirit to this method.

Another approach, used both by physicists and mathematicians is based
on the analytic properties of the S-matrix in the energy variable; see 
\cite{kn:LP}. Other approaches concentrated on the behavior of a reduced
Green's function, either by direct methods, or by studying its
analytic properties \cite{kn:Ho},\cite{kn:Or}.

The most commonly used approach is that of analytic dilation or, more
generally, 
 analytic  deformation \cite{kn:CFKS}, \cite{kn:HiSi}. 
  This method is very general, but
requires a choice of deformation group adapted to the problem at hand, as
well as technical analyticity conditions which do not appear to be necessary.
In this approach, the Hamiltonian of interest, $H$, is embedded in a
one-parameter family of unitarily equivalent operators, $H(\theta),\
\theta\in\R$. Under analytic
continuation in $\theta$ the continuous spectrum of $H$ is seen to move
and the eigenvalue, which was embedded in the continuum for the
unperturbed operator is now "uncovered" and isolated. Thus
Rayleigh-Schr\"odinger perturbation theory for an isolated eigenvalue
can be applied, and used to conclude that the embedded eigenvalue
generically perturbs to a resonance.
  The nonvanishing of the Fermi golden rule, (\ref{eq:fgr}),
   arises as a nondegeneracy
condition ensuring that we can see the motion of the embedded eigenvalue
at second order in perturbation theory. In our work, it arises as a
condition, ensuring the "damping" of states which are  
spectrally localized (with respect to $H$) about $\lambda_0$.
Analytic deformation techniques do not directly address the time
behavior, which require a separate argument 
 \cite{kn:GS},  \cite{kn:H}, \cite{kn:Skibsted}.  

Additionally, "thresholds" may not  be "uncovered" and therefore the
method of analytic deformation is unable to address the perturbation
theory of such points. Our time-dependent method can yield information about
thresholds, though it may be problematic to check the local decay
assumptions in intervals containing such points; see however example 5
in section 7, concerning the instability of a threshold eigenvalue of
$-\Delta + V(x)$.
 Finally,  in many cases,  previous approaches have required   
the potential to be dilation analytic, where  
 we only require $C^3$ behavior; see the concluding remarks of 
 appendix D for a discussion
of this point.

The paper is structured as follows. 
 In section 2 the mathematical
framework is explained and the main theorem (Theorem 2.1) is stated.
In section 3 the solution is decomposed relative to the
unperturbed operator, the key resonance is isolated and a dynamical
system of the form (\ref{eq:model}) is derived. Sections 4 and 5
contained the detailed estimates of the large time behavior of 
solutions. In section 6 we outline examples and applications. Sections
7-11 are appendices. Sections 7 (appendix A) 
 concerns the proof of the "singular"
local decay estimate of Proposition 2.1 and section 10 (appendix D), 
 some remarks on
a general approach to obtaining local decay estimates of the type assumed
in hypothesis (H4). In section 9  (appendix C) we present the details 
 of our expansion
of the complex frequency, $\omega_*$ (see ({\ref{eq:omegastar}) and
Proposition 3.3). In section 11 (appendix E) we give results on
boundedness of functions of self-adjoint operators in {\it weighted}
function spaces which may be of general interest.

\bigskip
\bigskip

\noindent{\bf Acknowledgements.\, }   We would like to thank
I.M. Sigal and T.C. Spencer 
for stimulating discussions. We also thank W. Hunziker and 
 J.B. Rauch for comments on the
manuscript. Part of this research was done  
 while MIW was on sabbatical
leave in the Program
in Applied and Computational Mathematics at Princeton University. MIW
would like to thank P. Holmes for hospitality and for a stimulating  
research environment. 
 This work is supported in part U.S. National Science
Foundation 
 grants DMS-9401777 (AS) and DMS-9500997 (MIW).

\section{ Mathematical framework and statement of the main
theorem} 
\setcounter{section}{2}
\medskip

In this section we first introduce certain necessary terminology and
notation. We then state the hypotheses {\bf (H)} and {\bf (W)} on
the unperturbed Hamiltonian, $H_0$, and on the perturbation, $W$. The
section then concludes with statements of the main results. 
\bigskip

For an operator, $L$, $||L||$, denotes its norm as an operator from  
$L^2$ to itself.
We interpret functions of a self adjoint operator as being 
 defined by the spectral theorem. In the special case where the 
the operator is $H_0$, we omit the argument, {\it i.e.} $g(H_0)=g$.

For an open interval $\Delta$, we denote an appropriate smoothed characteristic
function of $\Delta$ by $g_\Delta(\lambda)$. In particular, we shall take
$g_\Delta(\lambda)$ to be a nonnegative $C^\infty$ function, which is 
  equal to one on $\Delta$ 
 and zero outside a 
neighborhood of $\Delta$. The support of its derivative is furthermore
chosen to be small compared to the size of $\Delta$, {\it e.g.} less than
${1\over 10}|\Delta |$. We further require that $|g_\Delta^{(n)}(\lambda
)|\le c_n\ |\Delta |^{-n},\quad n\ge1$.

$P_0$ denotes the projection on $\psi_0$,\,  {\it i.e.} $P_0f\, =\,
(\psi_0,f)\psi_0.$

$P_{1b}$ denotes the spectral projection on 
 ${\cal H}_{pp}\cap \{\psi_0\}^\perp$, 
 the pure point spectral part of $H_0$ orthogonal
to $\psi_0$. That is, $P_{1b}$ projects onto the subspace of ${\cal H}$
spanned by all eigenstates other than $\psi_0$.  

In our treatment, a central role is played by the subset of the spectrum
 of the
operator $H_0$, ${\bf T^\#}$, on which a sufficiently rapid 
 local decay estimate holds. For a decay estimate to hold for
$e^{-iH_0t}$, one must
certainly project out the bound states  of $H_0$, but there may be other
obstructions to rapid decay. In scattering theory these are  
 called {\it threshold energies} \cite{kn:CFKS}. 
 Examples of thresholds are: (i) points of
stationary phase of a constant coefficient principle symbol 
 for two body Hamiltonians and (ii) 
 for $N-$ body Hamiltonians, zero and the eigenvalues of subsystems. 
We will not give  
a precise definition of thresholds.  For us it is sufficient to say
that away from thresholds the favourable local decay estimates for
$H_0$ hold.

Let $\Delta_*$ be union of intervals, disjoint from
$\Delta$, containing all thresholds of $H_0$, and a neighborhood of
 infinity.
\medskip
We then let
$$
P_1 = P_{1b} + g_{\Delta_*}
$$
where $g_{\Delta_*}\ =\ g_{\Delta_*}(H_0)$ is a smoothed characteristic
function of 
the set $\Delta_*$. 
We also define
\ba
\langle x\rangle^2 &=& 1+|x|^2,\nn\\
\overline Q &=& I-Q,\ {\rm and}\ \nn\\
P_c^\# &=& I-P_0 - P_1.
\label{eq:Pcsharp}
\ea
Thus, $P_c^\#$ is a smoothed out spectral projection of  
 the set ${\bf T^\#}$
defined as
\be
{\bf T^\#}= \sigma (H_0)\, -\, 
      \{\ {\rm eigenvalues,\  real\ neighborhoods\  of\  thresholds \ and\
 infinity}\ \} .
 \label{eq:Tsharp}
\ee
We expect $e^{-iH_0t}$ to satisfy good local decay estimates on the
range of $P_c^\#$; see 
 {\bf (H4)} below.

\bigskip

Next we state our {\bf hypotheses on $H_0$}.
\medskip

\noindent{\bf (H1)} $H_0$ is a self adjoint operator with dense domain
${\cal D}$, in $L^2(\R^n)$.

\noindent{\bf (H2)} $\lambda_0$ is a simple embedded eigenvalue of
$H_0$ with (normalized) eigenfunction $\psi_0$.

\noindent{\bf (H3)} There is an open interval $\Delta$ containing
$\lambda_0$ and no other eigenvalue of $H_0$.
\medskip

There exists $\sigma>0$ such that:

\noindent{\bf (H4)} {\it Local decay estimate:}
\nit
Let  $r \geq 2 +\varepsilon$ and
 $\varepsilon>0$.
If $\langle x\rangle^\sigma\ f\in L^2$ then 
\be
\|\langle x\rangle^{-\sigma} e^{-i H_0 t}P_c^\#  f\|_2 \le
\quad C \la t\ra^{-r}\ \|\langle x\rangle^\sigma\   f\|_2,
\label{eq:ld1}  
\ee

\noindent{\bf (H5)} By appropriate choice of a  real number $c$, the
 $L^2$ operator norm of 
 $\la x\ra^\sigma\ (H_0+c)^{-1}\ \la x\ra^{-\sigma}$
can be made sufficiently small.
\medskip

\nit{\bf Remarks:} 

\noindent (i) We have assumed that $\lambda_0$ is a simple eigenvalue
 to simplify the presentation. Our methods can be easily adapted to the
case of multiple eigenvalues.

\noindent (ii) 
 Note that $\Delta$ does not have to be small and  that $\Delta_*$ can be
chosen as necessary, depending on $H_0$.

\noindent (iii) 
In certain cases, the above local decay conditions can be proved
even when $\lambda_0$ is a threshold; see example 5 of
section 6.

\noindent (iv) 
 Regarding the verification of the local decay hypothesis, one approach
is to use  techniques based on the Mourre estimate \cite{kn:JMP},
 \cite{kn:SS}.
If $\Delta$ contains no threshold values then, quite generally,
  the bound (\ref{eq:ld1})
 hold with $r$ arbitrary and positive. See appendix D.
\medskip

We shall require the following consequence of hypothesis {\bf (H4)}.

\begin{prop}
  Let $r \geq 2 +\varepsilon$ and
 $\varepsilon>0$.  Assume $\mu\in {\bf T^\#}$. Then, for $t\ge0$   
\be
\|\langle x\rangle^{-\sigma} e^{-i H_0 t}(H_0-\mu -i0)^{-1}\
  P_c^\# f\|_2 \le
\ C \la t\ra^{-r+1}\ \|\langle x\rangle^\sigma  f\|_2,
\label{eq:ld3}
\ee
For $t<0$, estimate (\ref{eq:ld3}) holds with $-i0$ replaced by $+i0$.
\end{prop}
The proof is given in appendix A.
\bigskip

\nit We now specify the conditions we require of the perturbation, $W$.
\medskip

\noindent{\bf Conditions on $W$}
\medskip

\noindent{\bf (W1)} $W$ is symmetric and $H=H_0 + W$ is self-adjoint on
${\cal D}$ and there exists $c\in\R$ (which can be used in
{\bf (H5)}), such that $c$ lies in the resolvent sets of $H_0$ and $H$.
\medskip

\noindent{\bf (W2)} 

\nit For some $\sigma$, which can be chosen to be the
same as in {\bf (H4)} and {\bf (H5)}:
\ba
||| W |||\ &\equiv&\ 
 \|\langle x\rangle^{2\sigma}  W g_\Delta(H_0) \|\ +\
 \|\langle x \rangle^\sigma  W g_\Delta(H_0)  \langle x\rangle^\sigma\|\
   +\  \|\langle x \rangle^\sigma W(H_0 + c)^{-1}\ \langle
   x\rangle^{-\sigma}\ \|\ < \infty,\nn\\ 
   \label{eq:triplenorm}
   \ea
   and 
   \be \| \langle x \rangle^\sigma\ W\ (H_0+c)^{-1} \langle x
   \rangle^\sigma \|<\  \infty,\nn\ee
\medskip

\noindent{\bf (W3)} {\it Resonance condition\ -\ nonvanishing of the 
  Fermi golden rule:}
\be
\Gamma\equiv\ 
 \pi\  ( W\psi_0, \delta(H_0-\tilde\omega)(I-P_0)  W\psi_0)\neq 0
\label{eq:fgr}
\ee
for $\tilde\omega$ near $\lambda_0$, and  
\be
\Gamma \geq \delta_0 ||| W |||^2 \label{eq:fgrlb}
\ee
for some $\delta_0 > 0.$
\medskip

\noindent{\bf (W4)} $||| W ||| < \theta\ | \Delta |$
\medskip

\noindent for some $\theta>0$, sufficiently small, depending on the
properties of $H_0$, in particular the local decay constants, but not
on $|\Delta |$. 
\medskip

\nit {\bf Remark:} Let ${\cal F}^{H_0}_c$ denote the (generalized)
Fourier transform with respect to the continuous spectral part of
$H_0$. The resonance condition (\ref{eq:fgr}), can then be expressed as:
\be
\Gamma\ \equiv\ \pi\ \left|\ {\cal F}^{H_0}_c[W\psi_0](\lambda_0) \right|^2\
 >\ 0.
\label{eq:fgr-a} 
\ee
We can now state the main result:
\medskip
\medskip

\begin{theo}  Let $H_0$ satisfy the conditions {\bf (H)} and the
perturbation $W$ satisfy the  conditions {\bf (W)}.  Then 
\item{(a)} $H= H_0+W$ has no eigenvalues in $\Delta$.
\item{(b)} The spectrum of $H$ in $\Delta$ is purely absolutely
 continuous; in particular local decay estimates hold for $e^{-iHt}
 g_\Delta(H)$. Namely, for $\phi_0$ with $\la x\ra^\sigma\phi_0 \in L^2$, 
 as $t\to\pm\infty$
\be
||\ \la x\ra^{-\sigma}\ e^{-iHt}\ g_\Delta (H)\ \phi_0||_2\
=\ {\cal O}(\la t\ra^{-r+1}).
\ee

\item{(c)} For $\phi_0$ in the range of  $g_\Delta (H)$ we have (for
$t\ge0$)  
\ba
e^{-iHt}\phi_0 &=& 
\left(\ I\ +\ {\cal A}_W \right)
  \  \left(\ e^{-i\omega_*t}\ a(0)\ \psi_0 +
e^{-iH_0t}\ \phi_d(0)\ \right)\nn\\
   &+&\ \ {\cal R}(t).\label{eq:1}
\ea
Here, $ \|{\cal A}_W\|_{{\cal B}(L^2)}\le C|||W|||$, 
 $a(0)$ is a complex number and $\phi_d(0)$ is a  complex function
in the range of $P_c^\#$, which are determined by the initial data; see  
(\ref{eq:ansatz}-\ref{eq:orthog}). 

The complex frequency, $\omega_*$, is given by 
\ba
\omega_*\ 
&=&\ \omega\ -\ \Lambda\ -\ i\Gamma\ +\ {\cal O}(|||W|||^3), \ {\rm
where }\ 
\label{eq:omegastar}
\ea 
\ba
\omega\ &\equiv&\ \lambda_0\ +\ \left(\psi_0,W\psi_0\right),\label{eq:omegaeqn}
 \\ 
\Lambda\ &\equiv&\ 
 \left(\ W\psi_0,\ {\rm P.V.\ }\ (H_0-\omega )^{-1}\
 W\psi_0\ \right), \ {\rm and }\ 
\label{eq:Lambda}\\
 \Gamma &\equiv&\
  \pi\  ( W\psi_0, \delta(H_0-\omega)(I-P_0)  W\psi_0).
 \label{eq:Gamma}
 \ea

We also have the estimates:

\ba
||\la x\ra^{-\sigma}{\cal R}(t)||_2&\leq& C |||W|||,\quad t\ge0\label{eq:2}\\
||\la x\ra^{-\sigma}{\cal R}(t)||_2\ &\leq&\ 
							 C\ |||W|||^\varepsilon\ \la t\ra^{-r+1},
     \quad  t \ge |||W|||^{-2(1+\delta)},\ \delta>0,\
	 \varepsilon=\varepsilon(\delta)>0.\nn\\
	 \label{eq:4}
\ea
\end{theo}
\bigskip
\medskip
\nit{\bf Remark:} Though phrased in the setting of the space
$L^2(\R^n)$, our approach is quite general and our results hold with 
$L^2(\R^n)$ replaced by a Hilbert space, ${\cal H}$. In this general
setting, the
weight function, $\la x\ra$, is to be replaced by a "weighting
operator", $A$, in the hypotheses {\bf (H)}, {\bf (W)}
 and in the definition of
the norm of $W$, 
 $|||\ W\ |||$. Additionally, $P_c^\#$ can be taken to
be a smoothed out spectral projection onto the subspace of ${\cal H}$ where the
local decay estimate {\bf (H4)} holds. 

Given an eigenstate $\psi_0$ associated with an embedded eigenvalue,
$\lambda_0$, of the unperturbed  Hamiltonian, $H_0$, a quantity of
physical interest is the {\it lifetime} of the state $\psi_0$ for the
perturbed dynamics. To find the lifetime, consider the quantum
expectation value that the system is in the resonant state, $\psi_0$:
\be \left(\ \psi_0,\ e^{-itH}\ \psi_0\ \right).\ee 
Note that 
\be
e^{-iHt}\psi_0\ =\ e^{-iHt}\ g_\Delta(H)\psi_0\ +\ 
  e^{-iHt}\ \left( g_\Delta(H_0) - g_\Delta(H)\right) \psi_0.
\ee
 Theorem 2.1 and the techniques used in the proofs of Propositions 3.1 and 3.2 
  yield the following result
concerning the lifetime of the state $\psi_0$.
\begin{cor}
Let
\be 
H_*\ =\ H\ -\ {\rm Re\ } \omega_* \ I 
\ee
Then, for any $T>0$ there is a constant $C_T>0$ such that for 
 $0\le t\le T\ |||W|||^{-2}$ 
\be
\left|\ \left(\psi_0,e^{-iH_*t}\psi_0\right)\ -\ e^{-{\Gamma} \ t}\right|\
\le\  C_T|||W|||,\ {\rm \ as\ }\ |||W|||\to0.\nn
\ee
\end{cor}


\bigskip
\section{ Decomposition and isolation of resonant terms }
\medskip

We begin with the following decomposition of the solution of (\ref{eq:tdeps}):
\be
e^{-iHt}\phi_0 =\phi(t) = a(t)\psi_0 + \tilde\phi(t)\label{eq:ansatz}
\ee
\be
\left(\psi_0,\tilde\phi(t)\right) = 0 \qquad -\infty < t< +\infty.
\label{eq:orthog}
\ee
Substitution into (\ref{eq:tdeps}) yields 
\be
i\partial_t\tilde\phi = H_0\tilde\phi 
+ W\tilde\phi -(i\partial_t a -\lambda_0 a)\psi_0 
+ aW\psi_0 \label{eq:2.3}
\ee
Recall now that $I = P_0 + P_1 + P^\#_c$.  Taking the inner product 
of (\ref{eq:2.3}) with $\psi_0$ gives the amplitude equation:
\be
i\partial_t a = \left(\ \lambda_0 + (\psi_0,W\psi_0)\ \right)a 
+ (\psi_0 W P_1\tilde\phi) + (\psi_0,W\phi_d),\label{eq:2.4}
\ee
where, 
\be
\phi_d\  \equiv\  P^\#_c\tilde\phi.
\label{eq:phid}
\ee

The following equation for $\phi_d$ is 
obtained by applying $P^\#_c$ to equation (\ref{eq:2.3}):  
\be
i\partial_t\phi_d = H_0\phi_d + P_c^\# W\left( P_1\tilde\phi +
\phi_d\right) + a P_c^\# W\psi_0. \label{eq:2.5}
\ee

Our goal is to derive a closed system for $\phi_d(t)$ and $a(t)$. To
achieve this, we now propose to obtain an expression for 
$P_1{\tilde\phi}$, to be used in equations (\ref{eq:2.4}) and (\ref{eq:2.5}).
Since $g_\Delta (H)\phi(\cdot,t)=\phi(\cdot,t)$, we find
\be
(I-g_\Delta(H))\phi = (I-g_\Delta(H))\ [a\psi_0 + P_1\tilde\phi
+P_c^\#\tilde\phi]=0\label{eq:gbar}
\ee
or 
\be
 \left( I- g_\Delta(H)\gI\right)P_1\tilde\phi= -\overline
g_\Delta(H)\left[\ a\psi_0 + \phi_d\ \right],\label{eq:gbar1}
\ee
where $g_I(\lambda)$ is a smooth function, which is identically equal to
one on the support of $P_1(\lambda)$, and which has support disjoint
from $\Delta$. 
Therefore
\be
P_1\tilde\phi = -B\overline g_\Delta(H)\left(a\psi_0 + \phi_d\right),
 \label{eq:2.8}
\ee
where 
$$B = (I-g_\Delta(H)\gI)^{-1}.$$ 
This computation is justified by the following result which is proved in
appendix B.
\begin{prop}
 The operator $B=(I-g_\Delta(H)\gI)^{-1}$ is a bounded operator on
${\cal H}$
\end{prop}
\medskip

From  (\ref{eq:2.8}) we get 
\ba
\phi(t)\ &=&\  a(t)\psi_0+\phi_d+ P_1\tilde \phi\nn\\
         &=&\ \tg_\Delta(H)\left(
                     a(t)\psi_0 + \phi_d(t)\right),\label{eq:2.9}
\ea
with
\be
\tg_\Delta (H) \equiv I-B{\overline g}_\Delta (H)\ =\ B\ g_\Delta(H)\ 
(I-\gI).\label{eq:tildeg}
\ee
Although $\tg(H)$ is not really defined as 
 a function of $H$, we indulge in this
 mild abuse of notation to emphasize its dependence on $H$. In fact, we
 shall prove that, in some sense, $\tg_\Delta(H)\sim g_\Delta(H)\sim
 g_\Delta(H_0)$.

Substitution of the above expression (\ref{eq:2.8}) for $P_1\tilde\phi$ into
(\ref{eq:2.5}) gives: 
\be
i\partial_t\phi_d = H_0\phi_d + aP_c^\# W\tg_{\Delta}(H)\psi_0 +
P_c^\# W \tg_{\Delta}(H)\phi_d \label{eq:2.11}
\ee
and
\ba
i\partial_t a\  &=&\  [\lambda_0 + \left( \psi_0,W\tg_{\Delta}(H)\psi_0
\right)]a + \left( \psi_0,W\tg_\Delta(H)\phi_d\right)\nn\\
	&=&\ \omega\ a\ +\ (\omega_1 - \omega )\ a\ +\ \left(
	\psi_0,W\tg_\Delta(H)\phi_d\right),
\label{eq:2.12}
\ea
where
\ba
\omega\ &=&\ \lambda_0\ +\ \left(\psi_0,W\psi_0\right)\label{eq:omega},\\
\omega_1\ &=&\ \lambda_0\ +\ \left(
\psi_0,W\tg_{\Delta}(H)\psi_0\right).\label{eq:omega1}
\ea

The decay of $a(t)$ and $\phi_d$ is driven by a resonance. From equation
(\ref{eq:2.12}), the second term on the right hand side  of (\ref{eq:2.11})
oscillates approximately like $ e^{-i\lambda_0t}$. Since $\lambda_0$ lies in the
continuous spectrum of $H_0$, this term resonates with the  continuous
spectrum of $H_0$.
To make explicit the effect of this  resonance, we first write (\ref{eq:2.11})
 as an equivalent integral equation.
\ba
\phi_d(t) &=& e^{-iH_0 t}\phi_d(0) -i\int_0^t e^{-iH_0(t-s)}a(s)P_c^\#
W\tg_\Delta(H)\psi_0 ds\nn\\
 &&-i \int_0^t e^{-iH_0(t-s)}P_c^\# 
W\tg_\Delta(H)\phi_d ds\nn\\
&\equiv&  \phi_0(t) + \phi_{res}(t) + \phi_1(t). \label{eq:2.13}
\ea
\medskip

\nit{\bf Remark:} Using the above expansion and definitions, we have:
\ba
\phi(t)\ &=&\ e^{-i\omega_*t}a(0)\psi_0 + e^{-iH_0t}\left(
I-P_0\right)g_\Delta\phi_0\nn\\
&+&\  \left[ \tilde g_\Delta(H)-g_\Delta(H_0)\right] 
	 \left[ e^{-i\omega_*t}a(0)\psi_0 + e^{-iH_0t}
	 P_c^\#\phi_0\right]\ +\ {\cal R}(t),
	 \label{eq:exp1}\ea
where
\be
{\cal R}(t)\ =\ \tilde g(H)\left[\ \sum_{j=0}^4R_j(t)\psi_0\ +\ 
\phi_{res}(t)\ +\ \phi_1(t)\ \right].\label{eq:exp2}
\ee
The expansion in part (c) of Theorem 2.1 is obtained by 
estimates of the terms in (\ref{eq:exp1}) and (\ref{eq:exp2}).
These estimates are carried out in  sections 4 and 5.
\medskip

Our next goal is to obtain the leading order behavior of 
 $\phi_{res}(t)$. For $\epsilon >0$ introduce the following
regularization:

\be
\phi_{res}^\epsilon (t) = -i\int_0^t e^{-iH_0(t-s)}a(s) e^{\epsilon s}P^\#_c
W\tg_\Delta(H)\psi_0 ds.\label{eq:2.14}
\ee
Then, $\phi_{res}^\epsilon (t) \to \phi_{res}(t)$.  
 To extract the dominant oscillatory part of $a(t)$, we let 
\be
 A(t)\ = \ e^{i\omega t}\  a(t)
 \label{eq:2.16}
\ee

We now expand
$\phi_{res}^\epsilon (t)$ using integration by parts. 
\ba
\phi_{res}^\epsilon (t)&=& 
 -i\int_0^t e^{-iH_0t}e^{i(H_0-i\epsilon)s}a(s)P_c^\#W\tg_\Delta(H)
                                         \psi_0  ds \nn\\
 = &&-i\int_0^t e^{-iH_0t}e^{i(H_0-i\epsilon -\omega)s}[e^{i\omega
s}a(s)]P_c^\#W\tg_\Delta(H)\psi_0 ds \nn\\
 &\equiv& 
-i e^{-iH_0t}\int_0^t e^{i(H_0-i\epsilon -\omega)s}A(s)P_c^\#W\tg_\Delta(H)\psi_0 ds \nn\\
&=& - e^{-iH_0t}\ 
  \left[(H_0-\omega -i\epsilon)^{-1}e^{i(H_0-i\epsilon -\omega)s}
A(s)P_c^\# W\tg_\Delta(H)\psi_0 \right|_{s=0}^{s=t} \nn\\
+ &&e^{-iH_0t}\int_0^t (H_0-\omega -i\epsilon)^{-1}e^{i(H_0-i\epsilon
-\omega)s} \partial_t A(s) P_c^\#W\tg_\Delta(H)\psi_0 ds  
\label{eq:2.15}
\ea
With a view toward taking $\epsilon\downarrow 0$ we  first 
 note that by hypothesis {\bf (H)}, since $|||W|||$ is assumed
sufficiently small,  
 we have that $\omega\in\Delta$. The limit is therefore singular, and
 we'll find a resonant,
  purely imaginary, contribution coming from the boundary term at $s=t$.
Furthermore, to study the last term in (\ref{eq:2.15}) we will use the
equation:
\be
 \partial_t A = -ie^{i\omega t}(\psi_0,W\tg_\Delta(H)\phi_d)\ +\
 i(\omega - \omega_1) A.
\label{eq:2.17}
\ee
Now, taking  $\epsilon\to0$ we get, in $L^2(\la x\ra^{-2\sigma}\ dx)$, 
\medskip
\begin{prop}  The following expansion for
$\phi_{res}(t)$ holds:
\ba
 \phi_{res}(t) &=& -a(t)\ (H_0-\omega - i0)^{-1}\ P_c^\#\ W\tg_\Delta(H)
\ \psi_0 \nn\\
  &+&a(0)\ e^{-i H_0 t}\ (H_0-\omega - i0)^{-1}\ P_c^\#\ W\tg_\Delta(H)
\ \psi_0 \nn\\
 &-i&\ \int_0^t e^{-iH_0(t-s)}(H_0-\omega - i0)^{-1}
  P_c^\#W\tg_\Delta(H) \ \psi_0 \cdot \left(\psi_0,W\tg_\Delta(H)
\phi_d(s)\right)\ ds \nn\\
 &+i&\ (\omega - \omega_1)\ \int_0^t e^{-iH_0(t-s)}(H_0-\omega - i0)^{-1}
   P_c^\#W\tg_\Delta(H) \ \psi_0 \cdot a(s)\ ds\nn\\
&\equiv&  -a(t)(H_0-\omega - i0)^{-1}P_c^\#W\tg_\Delta (H) \psi_0 
  + \phi_2(t) + \phi_3(t) + \phi_4(t).
\label{eq:2.18}
\ea
\end{prop}
\medskip

\noindent{\bf Remark:}
To see that the terms in (3.21) are well defined
 we refer to the proof of Proposition 2.1 in appendix A.
 Localizing near and away from
 the energy $\omega$:
 \ba 
\left( H-\omega -i0\right)^{-1} e^{-iH_0t}P_c^\#\ &=&\nn\
\left( H-\omega -i0\right)^{-1} e^{-iH_0t} P_c^\#g_\Delta \ +\
\left( H-\omega -i0\right)^{-1} e^{-iH_0t} {\overline g}_\Delta\cr
&\equiv&\ T^t_{\Delta , 0}\ +\ S^t_{\Delta , 0}.
\nn\ea
	  In appendix A it is proved that for $\epsilon\ge0$,
	  $$T_{\Delta,\epsilon}^t,\ S_{\Delta,\epsilon}^t\ :  L^2(\langle
	  x\rangle^{2\sigma}\ dx) \mapsto
	  L^2(\langle
	  x\rangle^{-2\sigma}\ dx),\ \ t\ge0.$$

\bigskip

Substitution of (\ref{eq:2.13}) and (\ref{eq:2.18}) into
(\ref{eq:2.12}) yields the following equation for $a(t)$:

\be
 i\partial_t a(t) = \omega_* a(t)
+ \left(\psi_0,W\tg_\Delta(H)\{\phi_0(t)+\phi_1(t)+\phi_2(t)+\phi_3(t)
+\phi_4(t) \}\right).
\label{eq:2.19}
\ee
Here,
\ba
 \omega_* &=& \lambda_0 + \left(\psi_0, W\tg _\Delta(H) \psi_0\right) \nn\\
            & &- \left(\psi_0,W\tg_\Delta(H)  (H_0-\omega
-i0)^{-1}P_c^\#W\tg_\Delta (H) \psi_0\right) 
\label{eq:2.20}
\ea
In order see the resonant decay  we must first consider 
 the behavior of the {\it
complex} frequency $\omega_*$ for small $|||W|||$. The next proposition
contains an expression for $\omega_*$ which depends explicitly on the
"data" of the resonance problem, $H_0$ and $W$, plus a controllable
error.

\begin{prop}
\be
\omega_* = \lambda_0 + \left(\psi_0,W\psi_0\right)\ -\ \Lambda\ -\
i\Gamma\ +\ E(W),
\label{eq:2.21}
\ee
where
\ba
\Gamma\ &=&\ \pi \left(\ W\psi_0,\delta(H_0-\omega) (I-P_0)
 \  W\psi_0\right),\nn\\
\Lambda\ &=&\ \left(\  W\psi_0\ ,\ {\rm
P.V.}\ (H_0-\omega)^{-1}\ 
  W\psi_0\  \right),\nn \\
 E(W)\ &\le&\ C_1\ |||W|||^3,
\label{eq:2.22}
\ea
where $\omega$ is given by (\ref{eq:omegaeqn}). 
\end{prop}
The term, $\Gamma$, in (\ref{eq:2.22}) is the Fermi golden rule
appearing  in resonance hypothesis {\bf (W3)} ($\Gamma\ne0$).
\medskip

The proof of Proposition 3.3 is a lengthy computation which we 
present in appendix C.

We conclude this section with a summary of the coupled equations for
$\phi_d(t)$ and $a(t)$. 
\begin{prop} 
\ba
i\partial_t a &=& \omega_* a 
 + \left(\psi_0,W\tg_\Delta(H) \{\phi_0 + \phi_1 +\phi_2 +\phi_3 +\phi_4
 \}\right),
 \label{eq:2.23}\\
\phi_d(t) &=& e^{-iH_0 t}\phi_d(0) -i\int_0^t e^{-iH_0(t-s)}a(s)P_c^\#
W\tg_\Delta(H) \psi_0 ds \nn \\
 &&-i \int_0^t e^{-iH_0(t-s)}P_c^\#
W\tg_\Delta(H) \phi_d(s)\ ds, \label{eq:2.24}
\\
 &&{\rm where}\nn
\\   
\phi_0(t)&=& e^{-iH_0t}P_c^\#\phi_d(0) \label{eq:2.25}\\
    \phi_1(t)&=&-i\int_0^te^{-iH_0(t-s)}P_c^\#W\tg_\Delta(H) \phi_d(s)\ ds
\label{eq:2.26} \\
   \phi_2(t)&=& -a(0)e^{-iH_0t}\left(H_0 - \omega - i0\right)^{-1}
P_c^\#W\tg_\Delta (H)\psi_0 \label{eq:2.27}\\
\phi_3(t)&=&-i\int_0^te^{-iH_0(t-s)}(H_0-\omega -i0)^{-1}P_c^\#W\tilde
g_\Delta(H)\psi_0\cdot\left(\psi_0,W\tg_\Delta(H) \phi_d(s)\right)\ ds. 
 \\
 \label{eq:2.28}
\phi_4(t)&=& i (\omega - \omega_1)\ \int_0^t e^{-iH_0(t-s)}(H_0-\omega -
 i0)^{-1}
	P_c^\#W\tg_\Delta(H) \ \psi_0 \cdot a(s)\ ds
	\label{eq:phi4}
\ea
\end{prop}
To prove the main theorem we estimate $a(t)$ and $\phi_d(t)$ from
(\ref{eq:2.23}-\ref{eq:phi4}). 
Note that since ${\rm Im}\ \omega_*\ \sim\ -{\rm Im}\ \Gamma$ is negative, 
 it is evident that this resonant
contribution has the effect of
 driving $a(t)$ to zero.

\nit{\bf Remark:} Although we have the general result of Theorem 2.1, in
a given example it may prove beneficial to analyze the system
(\ref{eq:2.23}-\ref{eq:phi4}) directly in order to exploit special
structure.
\medskip

In the next two sections we estimate
 the solution over various time scales.

\section{ Local decay of solutions}

\medskip
 
In this section we begin our analysis of the large time behavior of
solutions. To prove local decay, we introduce the norms
$$
[a](T)\equiv \sup_{0\le s\le T}\langle s\rangle^{\alpha}
 | a(s)| \quad {\rm and }\quad [\phi_d]_{LD} (T)\equiv\sup_{0\le s\le
T}\langle s\rangle^{\alpha}||\langle x\rangle^{-\sigma}\phi_d(s)||_2,
$$
for which we seek to obtain upper bounds that are uniform in $T\in\R$.
Because of terms like $\phi_j(t),\ j=2,3,4$ (see Proposition 3.4)
and the singular local decay estimate of Proposition 2.1, it is natural
 study these norms with  
 $\alpha = r-1$.  
In this section, it turns out that we require the restriction  on
$\alpha$, $1<\alpha < 3/2$. Thus, throughout this section we shall
assume the constraints:
$$ \alpha\equiv r-1,\ \ 1<\alpha < 3/2.$$
In section 5 we relax the upper bound on $\alpha$.

\nit{\bf Remark: } 
In the estimates immediately below and in subsequent sections we shall
require bounds on the following quantities like: 
$\|\la x\ra^a\  W \tg_\Delta(H)\ \la x\ra^b  \|$ with  
$a,b\in \{0,\sigma\}$.
That all these can be controlled in terms of the norm $||| W |||$ is
ensured by the following proposition, which is proved in appendix B.
\begin{prop} 
For $a,b\in \{0,\sigma\}$, 
\be 
||\la x\ra^a W\tg_\Delta (H) \la x\ra^b ||\ \le\ C_{a,b}\ |||W|||
\label{eq:youcandoit}
\ee

\end{prop}

\medskip
We begin by estimating the local decay norm of
$\phi_d$. 
\bigskip

\centerline{\it Local decay estimates for 
$\phi_d(t)$}
\medskip

 From equation (\ref{eq:2.24})
\ba
 ||\langle x\rangle^{-\sigma}\phi_d(t)||_2 &\le&  
 ||\langle x \rangle^{-\sigma}
 e^{-iH_0t}\phi_d(0)||_2\   
 +\ \int_0^t\  
 |a(s)|\  ||\langle x\rangle^{-\sigma}
e^{-iH_0(t-s)}P_c^\#W\tg_\Delta(H) \psi_0||_2\ ds \nn \\ 
&+&\ \int_0^t
 ||\langle x \rangle^{-\sigma} e^{-iH_0(t-s)} P_c^\#W\tg_\Delta(H)
  \phi_d(s)||_2\ ds \nn\\
&\le&  C\la t\ra^{-r}||\la x\ra^\sigma\ \phi_d(0)||_2
+ C||\langle x\rangle^{\sigma} W\tg_\Delta(H) \psi_0||_2
\ \int_0^t\ \la t-s\ra^{-r}| a(s)| ds \nn\\
&&\ \ \ \ \  +\  ||\langle x\rangle^{\sigma}  W\tg_\Delta(H) \langle
x\rangle^{\sigma}||\ \int_0^t\ \la t-s\ra^{-r}\ 
 ||\langle x\rangle^{-\sigma}\phi_d(s)||_2\ ds 
\label{eq:3.1}
\ea
This implies, for $0\le t\le T$:
\ba
&&||\langle x\rangle^{-\sigma}\phi_d(t)||_2 \le\   
C\ \langle t\rangle^{-r}\ ||\la x\ra^{\sigma}\phi_d(0)||_2 \nn\\  
\ \  &+&\ C\langle
 t\rangle^{-\alpha}\left(\ ||\langle x\rangle^{\sigma} W\tilde
g_\Delta(H)\psi_0||_2 
\ [a](T)\ +\ 
 ||\langle x\rangle^{\sigma}  W\tilde
g_\Delta(H) \langle x\rangle^{\sigma}||
\ [\phi_d]_{LD} (T)\right)
\nn\\
\ &\le&\ C_1\ \la t\ra^{-r}\ 
  ||\la x\ra^{\sigma}\phi_d(0)||_2\ +\ C_2\ |||W|||\ \la t\ra^{-\alpha} 
 \left( [a](T)\ +\ [\phi_d]_{LD}(T)\ \right) . 
\label{eq:3.2}
\ea
It follows that
\be \
[\phi_d]_{LD} (T)\le C_1\ ||\la x\ra^{\sigma}\ \phi_d(0)||_2 +
 C_2 |||W|||\left(\ [a](T) + [\phi_d]_{LD} (T)\ \right),
\label{eq:3.3}
\ee
and therefore
\be \
\left(1-C_2\ |||W|||\right)\ [\phi_d]_{LD} (T)\le C_1\ ||\la x\ra^{\sigma}\ \phi_d(0)||_2 +
 C_2 |||W|||\ [a](T),
 \label{eq:3.3aa}
 \ee

An additional simple consequence of (\ref{eq:3.1}) and  the
orthogonality of the decomposition (\ref{eq:2.9}), is
\be
||\ \la x\ra^{-\sigma}\phi_d(t)\ ||_2\ \le\  C\la t\ra^{-r}
 \  ||\la x\ra^{\sigma}\phi_d(0)||_2\
 +\ C\ |||W|||\ ||\phi_0||_2.
\label{eq:3.3a}
\ee  
\vfil\eject 

\centerline{\it Estimation of $a(t)$}
\medskip

We estimate $a(t)$ using equation (\ref{eq:2.23}). This equation has
the form:
\be
 i\partial_t a = \omega_* a + \sum_{j=0}^4 F_j, \label{eq:3.4}
\ee
where
\be
F_j(t) \equiv \left(\psi_0,W\tg_\Delta(H) \phi_j\right).\label{eq:3.5}
\ee
Therefore,
\be
a(t) = e^{-i\omega_*t}a(0) + \sum_{j=0}^4 R_j(t),\label{eq:3.6}
\ee
where
\be
R_j(t) = -i\int_0^t e^{-i\omega_*(t-s)}F_j(s)\ ds. \label{eq:3.7}
\ee

We next estimate each $R_j$. In the course of carrying out the analysis we
shall frequently apply the following:
\begin{lem}
Let $\Gamma$, $\alpha$, and $\beta$ denote real numbers such that
$\Gamma>0$ and $\beta>1$. Define
\be
I_{\alpha,\beta}(t)\ =\  \la t\ra^\alpha\ \int_0^t e^{-\Gamma(t-s)}\la
s\ra^{-\beta}\ ds\ 
\nn\ee
Then, 
\noindent (i)
\be
 I_{\alpha,\beta}(t)\   
\le\ C\left(\ \la t\ra^\alpha\ e^{-{1\over2}\Gamma t}\ +\ \la
t\ra^{\alpha-\beta}\ \Gamma^{-1}\ \right).\nn
\ee
\noindent(ii) If $\alpha \le\beta$, we have 
\be
\sup_{t\ge0}I_{\alpha,\beta}(t)\ \le\ C\left(\ \Gamma^{-\alpha}\ +\
\Gamma^{-1}\ \right).\label{eq:4.13}
\ee
\end{lem}
\medskip

To prove this Lemma, note that
\ba
I_{\alpha,\beta}(t)\ &=&\ \la t\ra^\alpha\ 
 \left( \int_0^{t\over2}\ +\ \int_{t\over2}^t \right)\{\cdot\cdot\cdot\}\
 ds\nn\\
&\le&\ \la t\ra^\alpha\  e^{-{1\over2}\Gamma t}\ \int_0^{t\over2}
 \la s\ra^{-\beta}\ ds\ +\  C\la t\ra^{\alpha-\beta}\ \int_{t\over2}^t
e^{-\Gamma(t-s)}\ ds.\nn
\ea
Part (i) follows by explicitly carrying out the integrals, using that
$\beta>1$, and part (ii) follows by noting that the supremum over $t\ge0$
of the
expression obtained in (i). 
\medskip

\nit{\bf Estimation of $R_0(t)$:}
\be
 R_0(t) 
 = -i\int_0^t e^{-i\omega_*(t-s)}\left(\psi_0,W\tg_\Delta(H) e^{-iH_0s}P_c^\#
 \phi_d(0)\right)\ ds. \label{eq:3.8}
\ee
Estimation of the integrand gives:
\ba
&&\left|\left(\psi_0,W\tg_\Delta(H) e^{-iH_0s}\phi_d(0)\right)\right|
\nn \\
&= &\ | \left(\langle x\rangle^{\sigma} \tg_\Delta(H)
W\psi_0,\langle x\rangle^{-\sigma}e^{-iH_0s}\phi_d(0)\right)|\nn\\
&\le &\ ||\langle x\rangle^{\sigma} \tg_\Delta(H)
W\psi_0||_2\ ||\la x\ra^{-\sigma}e^{-iH_0s}\phi_d(0)||_2\nn\\
&\le &\ C\ ||\ \la x\ra^{\sigma} \tg_\Delta(H) W\psi_0||_2\ 
 ||\la x\ra^{\sigma}\ \phi_d(0)||_2\la s\ra^{-r}\nn\\
&\le&\ C\ |||W|||\ ||\la x\ra^{\sigma}\ \phi_d(0)||_2\ \la s\ra^{-r}.
\label{eq:3.9}
\ea
Use of (\ref{eq:3.9}) in (\ref{eq:3.8}) yields 
\be
| R_0(t)| \le C\ |||W|||\  
          ||\la x\ra^{\sigma}\ \phi_d(0)||_2\  
 \int_0^t\ e^{-\Gamma (t-s)}\ \la s\ra^{-r}\ ds
\label{eq:3.10}
\ee


\noindent
Multiplication of (\ref{eq:3.10}) by $\la t\ra^{\alpha}$,
  use of Lemma 4.1   
 and the lower bound for $\Gamma$, (\ref{eq:fgrlb}), yields the bound  
\be
 \langle t\rangle^{\alpha}\ |R_0(t)|\ \le\   
  C\ |||W|||^{1-2\alpha}\ ||\la x\ra^{\sigma}\ \phi_d(0)||_2,\ t\ge0.
 \label{eq:3.11}
\ee
It also follows from (\ref{eq:3.10}), since $r>1$, that 
\be
\left| R_0(t)\right| \le C\ |||W|||\ \|\la x\ra^\sigma\ \phi_d(0)\|_2.
\ee
\bigskip

\nit{\bf Estimation of  $R_1(t)$:}

We must bound the expression:
\be
 R_1(t) = -\int_0^t
e^{-i\omega_*(t-s)}\left(\psi_0,W\tilde
g_\Delta(H)\int_0^se^{-iH_0(s-\tau)}P_c^\#W\tg_\Delta(H)\phi_d(\tau)
\ d\tau\right)\ ds.
\label{eq:R1}
\ee

This can be rewritten as 
\be
R_1(t) = \int_0^t\ e^{-i\omega_*(t-s)}\ ds \left(\ \la x\ra^\sigma\tilde g_\Delta(H)W\psi_0,
			   \int_0^s \la x\ra^{-\sigma}e^{-iH_0(s-\tau)}P_c^\#W\tg_\Delta(H)
				\phi_d(\tau) \ d\tau\ \right)
\nn\ee
which satisfies the bound:
\be
\left| R_1(t)\right|\ \le
C\ ||\la x\ra^\sigma\tilde g_\Delta(H)W\psi_0 ||_2\  \int_0^t\ 
e^{-\Gamma 
(t-s)}\ ds \int_0^s 
	   ||\la x\ra^{-\sigma} e^{-iH_0(s-\tau )}P_c^\#W\tg_\Delta(H)\phi_d(\tau)||\ d\tau.
\nn\ee
Use of the assumed local decay estimate {\bf (H4)} gives that $R_1(t)$ is
bounded by 
\be
C\ ||\la x\ra^\sigma\tilde g_\Delta(H)W\psi_0 ||_2\ ||\la x\ra^\sigma W
\tilde g_\Delta(H) \la x\ra^\sigma ||\ \int_0^t\ e^{-\Gamma (t-s)}\ ds
\int_0^s\ \la s-\tau\ra^{-r} ||\la x\ra^{-\sigma}\phi_d(\tau)||_2\ d\tau
\label{eq:R1a}
\ee
and therefore
\be
 \left| R_1(t)\right|\ \le
C\ |||W|||^2\ \int_0^t\ e^{-\Gamma (t-s)}\ ds\int_0^s\ \la s-\tau\ra^{-r}
\la \tau \ra^{-\alpha}\ d\tau\ [\phi_d]_{LD}(T).
\nn
\ee
Using Lemma 4.1, we have  
\be 
\la t\ra^\alpha\   
\left| R_1(t)\right|\ \le\ C\ \left(\ 1\ +\ |||W|||^{2-2\alpha}\
\right)\ [\phi_d]_{LD}(T).
\label{eq:3.24a}
\ee 
\noindent
Furthermore, use of  (\ref{eq:3.3a}) in (\ref{eq:R1a})
gives:
\be
\left| R_1(t)\right|
 \le\ C\ |||W|||  
\ ||\la x\ra^{\sigma}\phi_0||_2\,\quad t\ge0.
\label{eq:3.25a}
\ee
\bigskip

\nit{\bf  Estimation of $R_2(t)$:}
\ba
&& R_2(t)\, = \nn\\
      && ia(0)\int_0^t e^{-i\omega_*(t-s)}
\left(\psi_0,W\tg_\Delta(H) e^{-iH_0s}\left(H_0-\omega-i0\right)^{-1} 
                          P_c^\#
          W\tg_\Delta(H)\psi_0\right)\ ds. 
\nn\\
\label{eq:3.26}
\ea
Therefore, by Proposition 2.1,
\be
\left| R_2(t)\right|\ \le\ 
 C\ |a(0)|\ |||W|||^2\  \int_0^t e^{-\Gamma (t-s)}\la s\ra^{-r+1}\ ds.
\label{eq:3.27}
\ee
A first simple consequence, since $r>2$, is that 
\be \left| R_2(t)\right|\ \le\ C\ |a(0)|\ |||W|||^2.
\nn
\ee
Next, multiplication of (\ref{eq:3.27})  by $\la t\ra^{\alpha}$, 
 taking supremum over the interval 
  $0\le t\le T$ and applying Lemma 4.1 yields the
bound:
\be
\langle t\rangle^{\alpha}\left|R_2(t)\right|\ 
 \le\ C_1\ |a(0)|\left(\ 1  + |||W|||^{2-2\alpha}\ \right),
\label{eq:3.31}
\ee
\medskip
\medskip

\nit{\bf Estimation of $R_3(t)$:}
We begin by recalling:
\be
R_3(t)\ =\ -i\int_0^t\ e^{-i\omega_*(t-s)} F_3(s)\ ds. \label{eq:R3}
\ee
Therefore,
\be
\left| R_3(t)\right|\ \le\ C\ \int_0^te^{-\Gamma(t-s)}\ |F_3(s)|\
ds.\label{eq:R3abs}
\ee
$F_3(t)\ =\ \left(\psi_0,W\tg_\Delta(H)\phi_3(t)\right)$ is given
explicitly by the expression:
\ba
&{}& -i\ \int_0^s\ d\tau 
\left(\psi_0, W\tg_\Delta(H)  e^{-iH_0(s-\tau)} 
\left(H_0-\omega-i0\right)^{-1}P_c^\#W\tg_\Delta(H)\psi_0\right)
\times \left(\psi_0,W\tg_\Delta(H)\phi_d(\tau)\right)\nn\\
&=&\ -i\ \int_0^s\ d\tau
\left(\la x\ra^\sigma \tg_\Delta(H) W\psi_0,\ \la x\ra^{-\sigma}\
 e^{-iH_0(s-\tau)} 
\left(H_0-\omega-i0\right)^{-1}P_c^\#\la x\ra^{-\sigma}\cdot  
 \la x\ra^\sigma W\tg_\Delta(H)\psi_0\right)\nn\\
&&{}\quad\times \left(\la x\ra^\sigma\tg_\Delta(H) W\psi_0,\ \la
x\ra^{-\sigma} \phi_d(\tau)\right)\nn\\
\label{eq:3.32}
\ea
Estimation of $F_3(t)$ yields the bound:
\be\left| F_3(t)\right|\ \le\  C_W\ 
\int_0^s\ ||\langle x\rangle^{-\sigma}
 e^{-iH_0(t-\tau)}\ (H_0-\omega - i0)^{-1}
  P_c^\#\langle x\rangle^{-\sigma}||\ ||\la x\ra^{-\sigma}
   \phi_d(\tau)||_2\ d\tau,\nn\\
   \label{eq:3.33}
\ee
   where
\be
C_W\ =\ 
||\langle x\rangle^{\sigma} \tg_\Delta(H) W\psi_0||_2^2\ 
 \cdot ||\langle x\rangle^{\sigma} W\tg_\Delta(H)\psi_0||_2\ 
 \le\ C\ |||W|||^3
\label{eq:3.33a}
\ee
By Proposition 2.1 and (\ref{eq:3.33}-\ref{eq:3.33a}),
\be
\left| F_3(s)\right|\ \le\ C\ |||W|||^3\ \int_0^s \la s-\tau\ra^{-r+1}
\|\la x\ra^{-\sigma} \phi_d(\tau)||_2\ d\tau 
\label{eq:F3est}
\ee
If we bound $||\la x\ra^{-\sigma}\phi_d(\tau)||_2$ simply by
$||\phi_0||_2$ we obtain from (\ref{eq:F3est}) and
(\ref{eq:R3abs}):
\be
\left| R_3(t)\right|\ \le\ C\ |||W|||\ ||\phi_0||_2.
\label{eq:3.37a}
\ee
On the other hand, bounding $||\la x\ra^{-\sigma}\phi_d(\tau)||_2$ by 
$[\phi_d]_{LD}(T)\ \la \tau\ra^{-\alpha}$ ($\alpha=r-1$) in
(\ref{eq:F3est}) we obtain:
\be
\left| F_3(s)\right|\ \le\ C\ |||W|||^3\ \la s\ra^{-\alpha}\
[\phi_d]_{LD}(T).
\label{eq:F3this}
\ee
Finally, use  of (\ref{eq:F3this})  in (\ref{eq:F3est}) and applying Lemma 4.1  
  we have 
\be
\la t\ra^{\alpha}\ 
  \left| R_3(t)\right|\ \le\ C\ \
 \left(\ |||W|||\ +\ |||W|||^{3-2\alpha}\ \right)\ [\phi_d]_{LD}(T).
\label{eq:R3est}
\ee

\medskip

{\bf Estimation of $R_4(t)$:}

\ba R_4(t)\ &=&
    -i\int_0^t\ e^{-i\omega_*(t-s)}\ 
	  \left(\psi_0,W\tg_\Delta(H)\phi_4(s)\right)\ ds\nn\\
	&=&\ (\omega -\omega_1)\ \int_0^t\ e^{-i\omega_*(t-s)}\ 
	 \left(\tg_\Delta(H)W\psi_0,\int_0^s a(\tau ) e^{-iH_0(s-\tau)}
	  P_c^\#W\tg_\Delta(H)\psi_0\right)\ d\tau
	  \nn\ea
By Proposition 2.1,
\ba
\left| R_4(t)\right|\ &\le& |\omega -\omega_1 |\ |||W|||^2\ \int_0^t\
e^{-\Gamma (t-s)}\ ds\ \int_0^s \la s-\tau\ra^{-r+1}\ |a(\tau)|\ d\tau \nn\\
&\le& |\omega -\omega_1 |\ |||W|||^2\ \int_0^t e^{-\Gamma (t-s)} \la
s\ra^{-\alpha}\ ds\ [a](t).
\nn
\ea
We now estimate the $|\omega -\omega_1 |$. By
(\ref{eq:omega}-\ref{eq:omega1}),
\be
\omega_1 -\omega\ =\ \left(\psi_0,W\tg_\Delta(H)\psi_0\right)\ - 
					  \left(\psi_0,W\psi_0\right)\ \equiv\ \beta.
\nn\ee
An explicit expression, (\ref{eq:beta}), 
 is obtained for $\beta$ in appendix C:
 \be
 \beta = - \left( W\psi_0, \; B\bar{g}_\Delta(H)(H - \lambda_0)^{-1}\;
 W\psi_0 \right) .
 \label{eq:beta1}
 \ee
\bigskip
From Theorem 11.1 of appendix E and an argument along the lines of the
proof of (\ref{eq:youcandoit}) we have $|\beta |\le C|||W|||^2$. 
Therefore, using Lemma 4.1, we find 
\be
\la t\ra^\alpha \left| R_4(t)\right|\le C\ |||W|||^{4-2\alpha}\ [a](T).
\label{eq:R4est}
\ee
If $\alpha <3/2$, then
\be
\la t\ra^\alpha\ \left| R_4(t) \right|\ \le\ C\ |||W|||\ [a](T).
\nn
\ee
\bigskip

\centerline{\it Closing the estimates and completion of the proof}

We can now combine the upper bounds (\ref{eq:3.11}), (\ref{eq:3.24a}), 
 (\ref{eq:3.31}), (\ref{eq:R3est}) and (\ref{eq:R4est})  
for the $R_j(t), 0\le j\le4$ to obtain, via (\ref{eq:3.6}), 
the following upper bound for $a(t)$ provided $|||W||| < 1/2$:
\ba
 [a](T)\ &\le&
  c_1\ |a(0)|\ |||W|||^{-2\alpha}\ +\  
  c_2\ |||W|||^{1-2\alpha}\|\la x\ra^\sigma\ \phi_0\|_2\nn\\
\quad\ &+&\ c_3\ \left(\ 1\ +\ |||W|||^{2-2\alpha}\ \right)\
  [\phi_d]_{LD}(T). 
\label{eq:3.41}
\ea

Substitution of this bound into  (\ref{eq:3.3}) gives the following
bound for $\phi_d$: 
\ba
[\phi_d]_{LD}(T)\  &\le&\  C_0\ \left(\ 1\ +\ |||W|||^{2-2\alpha}\ \right)\ 
 \|\ \la x\ra^{\sigma}\ \phi_d(0)\ \|_2\ +\ 
 c_1\ |||W|||^{1-2\alpha}\ |a(0)|\nn\\
 &+&\ C_3\ \left(\ |||W|||\ +\ 
 |||W|||^{3-2\alpha}\ \right)\  [\phi_d]_{LD}(T).
\label{eq:3.42a}
\ea
Use of (\ref{eq:3.42a}) as a bound for the last term in (\ref{eq:3.41})
yields a bound for $[a](T)$:
\be
[a](T)\ \le\ c_1\ |a(0)|\ |||W|||^{-2\alpha}\ +\ c_2\ 
 |||W|||^{1-2\alpha}\ ||\la x\ra^\sigma\
\phi_d(0)||_2.
\label{eq:3.42b}
\ee
Finally, 
 for $|||W|||$ sufficiently small and $\alpha<3/2$ we have:
\be
[\phi_d]_{LD}(T)\ \le\ C\ \left(\ 1\ +
 \ |||W|||^{2-2\alpha}\ \right)\ ||\la x\ra^\sigma\ \phi_d(0) ||_2
 \ +\ C\ |||W|||^{1-2\alpha}\ |a(0)|.  
\label{eq:3.42c}
\ee
Taking $T\to\infty$ we conclude the decay of $\phi(t)$, with initial data 
$\phi_0$ in the range of $P_\Delta(H)$, with rate
$\langle t\rangle^{-\alpha}, 0\ <\ \alpha\  <\ 3/2$. 
 It follows \cite{kn:RS4} 
 that the
interval $\Delta$ consists of absolutely continuous spectrum of $H$, as
asserted in parts (a) and (b) of Theorem 2.1.
\medskip

\medskip

\bigskip
\section{Local decay of solutions for  large $r$} 

In the preceding subsection we proved the decay of solutions,
$\phi(t,x)$, in the local decay sense, with a slow rate of decay $\la
t\ra^{-\alpha}$ with $1<\ \alpha\ <\ 3/2$;\ $\alpha=r-1$.
A consequence of this result is that, in the interval $\Delta$, the
spectrum of $H$ is 
 absolutely continuous. Now if $\Delta$
contains no thresholds of $H$, we expect decay as
$t\to\infty$ at a rate which is faster than any polynomial. (For
example, this is what one has for constant coefficient dispersive
equations for energy intervals containing no points of stationary phase.) 
In this section we
show that this result holds in the sense of (\ref{eq:4}) in Theorem 2.1. 
 This requires some adaptation of the methods of section 4.
 We  shall indicate here only 
 the required modifications to the
argument of the previous section. 

\nit (1)  The origin of the restriction $\alpha\ <\ 3/2$ can be traced to
our application of part (ii) of Lemma 4.1. In particular,  
 in obtaining (\ref{eq:4.13}) we use that: 
\be
\sup_{t\ge0}\langle t\rangle^{\alpha}e^{-\Gamma t}\ =\ {\cal
O}(\Gamma^{-\alpha})
\label{eq:3.45}
\ee
 It follows that certain coefficients are found to be large for $|||W|||$
small, an obstruction to closing the system of estimates for $[a]$ and
$[\phi_d]_{LD}$, unless $\alpha\ <3/2$. This is remedied by taking the supremum in (\ref{eq:3.45})
 over
$t$ in the interval $[\Gamma^{-1-\delta}, T]$, where $\delta>0$.
\begin{lem}
Let $M\equiv \Gamma^{-1-\delta}\sim |||W|||^{-2(1+\delta)}$; see {\bf (W3)}.
There exists $\theta_*>0$ such that if $||| W |||<\theta_*$ and $t\ge M
$, then 

\noindent
(a) 
$\la t\ra^{r-1}\int_0^t e^{-\Gamma(t-s)}\la s\ra^{-r}\ ds\ 
\le\ C\ \Gamma^\delta\ \sim\ |||W|||^{2\delta}.$

\noindent (b)
 $\la t\ra^{r-1}\int_0^t e^{-\Gamma(t-s)}\la s\ra^{-\alpha}\ ds\ 
 \le\ C\ \Gamma^{-1} .$
\end{lem}
\medskip

\noindent (2) Assume $r>2$, ($\alpha>1$). 
 The analysis of section 4
  yields a coupled system of integral
inequalities for the functions $a(t)$ and 
\be
L(t)\ \equiv\ \|\la x\ra^{-\sigma}\phi_d(t)\|_2.\nn\\ 
\ee
The precise form of these inequalities can be seen as follows. Let 
\be I_r\{L\}(t)\ =\ \int_0^t\ \la t-s\ra^{-r}\ L(s)\ ds.\nn
\ee
Then, by  (\ref{eq:3.1}), (\ref{eq:3.6}) and the estimates for $R_j(t),\  
j=0,1,2,3,4,$,  the inequalities for $L(t)$ and $a(t)$ take the form:
\vfil\eject 

\ba
L(t)\ &\le&\  C_0\ \la t\ra^{-r}\ +\ C_1\ |||W|||\ I_r\{|a|\}(t)\ +\ 
C_2\ |||W|||\ I_r\{L\}(t)\nn\\
|a(t)|\ &\le&\ A_0e^{-\Gamma t}\ +\ A_1\ |||W|||\ I_r\{e^{-\Gamma s}\}(t)\ +\ 
 A_2\ |||W|||^2\ I_{r-1}\{e^{-\Gamma s}\}(t)\nn\\
 &+&\ A_3\ |||W|||^2\ \int_0^te^{-\Gamma(t-s)}\ I_r\{L\}(s)\ ds\ +\ 
 A_4\ |||W|||^3\ \int_0^te^{-\Gamma(t-s)}I_{r-1}\{L\}(s)\ ds\nn\\
 &+&\ 
  A_5\ |||W|||^4\ \int_0^te^{-\Gamma(t-s)}I_{r-1}\{|a|\}(s)\ ds,
 \nn\\
 \label{eq:Laeqns}
 \ea
 where the $C_j$ and $A_j$ denote positive constants.

\noindent
(3) The procedure is first to consider the functions $L(t)$ and $a(t)$ on a
large, but finite time interval: $0\le t\le \Gamma^{-1-\delta}\equiv M$, where
$\delta$ is positive and suitably chosen. An explicit bound for $L(t)$ and
$a(t)$ can be found by iteration of the inequalities (\ref{eq:Laeqns}).
For this, we use the following estimate of $I_r\{e^{-\Gamma s}\}$, which is
proved using integration by parts:
\be
I_r\{e^{-\Gamma s}\}\ \le\ c_0e^{-\Gamma t}\ +\ 
\sum_{k=1}^{r-2}\ c_k\ \Gamma^{k-1}\ \la t\ra^{-r+k-1} 
 \ +\ c_{r-1}\ \Gamma^{r-2-\rho}\ \la t\ra^{-1},
 \ee
 where $\rho>0$ is arbitrary.

 \noindent
 (4) To show decay for arbitrary, in particular large,  $\alpha = r-1$, and
 the estimates of ${\cal R}(t)$ of Theorem 2.1, we introduce the norms:
\be
   [a]^{\Gamma}(T) \equiv
  \sup_{M\le t\le T}\   
 \langle t\rangle^{\alpha}|a(t)|\label{eq:Gammanormofa}
\ee
and
\be
 [\phi_d ]_{LD}^{\Gamma}(T) \equiv
  \sup_{M\le t\ \le\ T } \  
  \la t\ra^{\alpha} ||\la x\ra^{-\sigma}\phi_d(t)||.
\label{eq:Gammanormofphi}
\ee
We now reexpress the system (\ref{eq:Laeqns}) for $L(t)$ and $a(t)$
by breaking the time integrals in  (\ref{eq:Laeqns}) into a part over the
interval $[0,M]$ and a part over the interval $[M,t]$. Using the estimate of
part (3) above, the integrals over $[0,M]$  are estimated to be of order
$|||W|||^\varepsilon\ \la t\ra^{-r+1}$ for some $\varepsilon=\varepsilon(\delta)$. 

In this way, the resulting system for $L(t)$ and $a(t)$ is now reduced to
one which can be studied using Lemma 5.1 and the approach of section 4.
Using this approach estimates for the norms (\ref{eq:Gammanormofa}) and 
(\ref{eq:Gammanormofphi}), and consequently of 
${\cal R}(t)$ can be obtained.

\section{ Examples and applications}

In this section we sketch examples and  applications of  
 Theorem 2.1. Most of these examples have been previously studied, under
 more stringent hypotheses on $H_0$ and $W$, 
 {\it e.g.} some type of analyticity: 
dilation analyticity for the Helium atom,
translation analyticity for the Stark Hamiltonian; see
 \cite{kn:CFKS} and references cited therein.  Theorem 2.1 enables
us to relax this requirement and gives the detailed time-behavior of
solutions near the resonant energy at all time-scales. 
  Example 5 concerns the instability of an
eigenvalue embedded at a threshold, a result which we believe is new and
not tractable by techniques of dilation analyticity.
\bigskip

We begin with the remark that in the examples below, one can often replace
the operator, $-\Delta$ by 
 $H_1\equiv\omega (p)$, where $p=-i\nabla$. 
The necessary hypothesis on local decay, {\bf (H4)}, is reduced to  
its verification for $H_1 + V$. By the general discussion of local decay
estimates of appendix D (see also \cite{kn:S3}), we have:

\begin{theo}

 The operator $H = H_1 + V$ 
 satisfies the required local
decay estimates of {\bf (H4)} under the following hypotheses: 
\bigskip

\nit Hypotheses on $\omega(p)$:

\nit (i)  $\omega(p)$ is real valued and $\omega(p)\to \infty$ as
$|p|\to\infty$.
\medskip

\nit (ii) $\omega(p)$ is $C^m$ function, $m\geq 4$.
\medskip

\nit (iii) $\nabla_p\omega =0$ on at most finitely many points, in any
compact domain.
\bigskip

\nit Hypotheses on $V(x)$:

 $V(x)$ is real valued and such that 
\medskip

\nit (V1) $V(x),\ x\cdot\nabla V,\ (x\cdot \nabla)^2 V,\  (x\cdot \nabla)^3 V$
are all $g(H_1)$ bounded for $g\in C_0^\infty$.
\medskip

\nit (V2) $|V(x)| = 0(\langle x\rangle^{-\epsilon}), \epsilon > 0, |x|\to
\infty$.
\medskip
 
\nit (V3)  $\chi_R\ V, \chi_R\ (x\cdot \nabla)^m\ V, m=1,2,3$ are
$g(H_1)$-compact, for $\chi_R\equiv \chi_{[R,\infty)}(|x|)$ with
some $R>0$.
\end{theo}
\bigskip

The proof of this result follows from the procedure outlined in appendix
D where we use the hypotheses on $\omega(p)$ and $V$ and the choice for
 the
operator $A$ is:
$$
A={1\over 2}(x\cdot\nabla_p\omega + \nabla_p \omega\cdot x).
$$

\medskip
\nit {\bf Remark:}  
 Due to lack of assumptions on analyticity of $\omega(p)$ or
$V(x)$ one cannot simply apply the technique of analytic deformation used
in other approaches.
\medskip

\noindent{\bf Example 1:} - {\it Dispersive Hamiltonian}

  With the above assumptions on $\omega(p)$ and $V(x)$, 
 Theorem 2.1 applies directly to the operator $H_0\equiv\omega(p) + V(x)$.
\medskip

\noindent{\bf Example 2} - {\it Direct Sum}

Let
$$
H_0= \pmatrix{
-\Delta_x & 0\cr
0 & -\Delta_x \ +\ q(x)}
$$
acting on $\C^2 \otimes L^2(\R^n)$, where $q(x)$ is a
 well behaved potential having some positive discrete eigenvalues.
 An example of this type is considered in \cite{kn:Waxler95}.

Consider, for example, the case where $q(x)\ =\ P(x)$, is a polynomial which is bounded below. In
this case, the spectrum of  $-\Delta_x +P(x)$  is discrete and consists
of  an infinite set of eigenvalues
 $\lambda_1 <
\lambda_2\cdots$ with corresponding 
   eigenfunctions $\psi_1, \psi_2,\cdots$ .  The
spectrum of $H$ is then 
$$
 \{\  {\rm\  eigenvalues\  of\  }\  -\Delta_x + P(x)\} \cup [0,\infty)
$$
and therefore $H_0$ has  nonnegative eigenvalues embedded in its 
 continuous spectrum.

Let 
$$
W=\pmatrix{ 0 & W(x)\cr
W(x) & 0}
$$
with $W$ satisfying conditions ${\bf (W)}$.

\begin{theo}
For $H_0$ and $W$ as above, if for some strictly postive simple 
 eigenvalue $\lambda >0$
 the resonance condition (Fermi golden rule) (\ref{eq:fgr}) holds, then in an 
 interval $\Delta$ around $\lambda$, 
the spectrum of $H$ is absolutely 
 continuous and the other conclusions of Theorem 2.1 hold.
  Furthermore, if $n > 4$, Theorem 2.1 holds even when  $\lambda = 0$
is an eigenvalue.
\end{theo}
\nit{\it proof:}
In this case local decay must be proved for $-\Delta_x$, with
$r>2$.  This is well known.
What is more, if the spatial dimension is larger than four, $n>4$, 
 then $\lambda=0$ is also
allowed, since in this case we use
\ba
||\langle x\rangle^{-{n\over 2}-\epsilon} e^{i\Delta_x t} \psi ||_2\ &\leq& 
||\ \langle x\rangle^{-{n\over 2} - \epsilon}\ ||_2\  
                        || e^{i\Delta_x t}\psi ||_\infty
\nn\\
&\leq& \ C\  t^{-n/2}\  ||\psi ||_1.
\nn
\ea
Hence, for $\psi\in {\cal D}(\langle x\rangle^{n/2 +\epsilon})$, and $n>4$
we have the necessary decay:
$$
||\langle x\rangle^{-{n\over 2} -\epsilon} e^{i\Delta_x t} \langle
x\rangle^{-{n\over 2} - \epsilon} || \leq Ct^{-n/2}
$$
with $r\ =\ {n\over 2}\ > 2$.  
\bigskip

\noindent{\bf Example 3} - {\it Tensor Products}
\medskip

Let $H_0 = {\bf 1}\otimes\ h_1 + h_2 \otimes  {\bf 1}$
acts on $L^2(\R^n_{x_1})\otimes L^2(\R^n_{x_2}),$
where 
\be 
h_1\ =\ -\Delta_{x_1},\ {\rm and}\ h_2\ =\ -\Delta_{x_2}+q(x_2).
\label{eq:tenham}
\ee
Then, 
\be
\sigma(H_0)\ =\ 
\{\lambda\  :\ \lambda=\lambda_1+\lambda_2,\ \lambda_1\in\sigma
(-\Delta_{x_1})\ {\rm and}\ \lambda_2\in\sigma (-\Delta_{x_2} + q(x_2))\
 \}.\nn
\ee 

Let $W(x_1,x_2)$ act on $L^2\otimes L^2$, satisfying ${\bf (W)}$, 
 with $\langle
x\rangle^2\equiv 1+|x_1|^2+|x_2|^2$.

Then we have
\begin{theo} 

The embedded eigenvalues of $H_0$ are unstable and
Theorem 2.1 holds.
\end{theo}

\noindent{\bf Example 4}
\ {\it Helium type Hamiltonians}\ -\  \cite{kn:RS4}
\medskip

Consider $H_0$ as in example 3 with:
\ba
&&h_1\ =\ -\Delta_{x_1}-|x_1|^{-1},\quad \ h_2 =\
-\Delta_{x_2}-|x_2|^{-1}.\ 
\label{eq:helium}
\ea
Also, let $W$ be of the form:
$$
W(x_1,x_2) = W(x_1-x_2).
$$
In this case the weight $\langle x\rangle^2 = 1 + |x_1|^2 + |x_2|^2$.  
We now discuss the hypothesis {\bf (W)}.

 $H_0$ has infinitely many negative eigenvalues embedded in the
continuous spectrum \cite{kn:CFKS}. 
  If $\Delta$ is a subinterval of the negative real line containing exactly one
negative eigenvalue, $E$, then $g_\Delta$ is a sum of terms of the form:
\be
g^c_{\Delta-E}(h_1)\otimes {\bf P}\ {\rm and }\ 
 {\bf P}\otimes g^c_{\Delta -E}(h_2).\
\ee
Here, $g^c_{\Delta - E}(h_j)$ a  spectral projection onto the continuous
spectral part associated with an interval $\Delta - E$, the translate of
$\Delta$ by $-E$, and ${\bf P}$ denotes a
(negative) bound state projection. Thus, $g_\Delta$ localizes either the
$x_1$ or the $x_2$ variable and so while  
$\langle x\rangle^{2\sigma} W$ is not bounded, 
we do have that 
$$
\langle x\rangle^{2\sigma} Wg_\Delta(H_0)
$$ 
is bounded provided, for example, $W$ is short range.
\medskip

In the case, where $W$ is long range, {\it i.e.} 
\be 
W(x_1-x_2)\ =\ {\cal O}(\la x_1-x_2\ra^{-1})
\nn\ee
we first prove a {\it minimal velocity bound} and then use it to get 
local decay.

Going back to (\ref{eq:2.24}) we estimate:
\be
\|\ F\left({|x|\over t}\le\eta\right)\ \phi_d(t)\ \|_2
\nn
\ee
using the known propagation and minimal velocity estimates for $H_0$
 \cite{kn:SS}. The problematic term, which is the last term on the
right hand side of (\ref{eq:2.24}) is then bounded by:
\ba
&&c_1\ \int_0^t\ \la t-s\ra^{-1-\varepsilon}\ \| \la x\ra^{1/2+\delta}\ 
 W\ 
\tilde g_\Delta (H) \|_2\ \| F\left({|x|\over s}\le\eta\right) 
 \phi_d(s)\|_2\ ds\nn\\
&&+\ c_2\ \int_0^t\ \la t-s\ra^{-1-\varepsilon}\ 
\| \la x\ra^{1/2+\delta}\ W\ \tilde g_\Delta (H)\   
F\left({|x|\over s}\ge\eta\right)\ \|\ \|\phi_d(s)\|_2\ ds.
\nn
\ea
Since 
\be 
\| \la x\ra^{1/2+\delta}\ W\ \tilde g_\Delta (H)\   
F\left({|x|\over s}\ge\eta\right)\ \|\ \le\ 
 c_3\ |||W|||\ \la s\ra^{-1/2 +\delta},
\nn
\ee
we can close the inequalities and obtain
\be
\la t\ra^{1/2 -\delta}\ \|\ F(\left({|x|\over t}\le\eta\right)\ 
 \phi_d(t)\ \|\ 
 \le\ c_0\ +\ c_1\ \sup_{0\le s\le t} |\la s\ra^{1/2-\delta}\ a(s)|.
\ee

The above estimate, together with the estimates for $a(t)$ lead to local
decay with a rate $\la t\ra^{-1/2 +\delta}$. This rate is not sufficient
to preclude singular continuous spectrum. However, the Mourre estimate
holds in this interval for $H_0+W$ which implies local decay and absence
of singular continuous spectrum; see Theorem 10.1 and Theorem 10.2 in
appendix D.
\medskip

\noindent{\bf Example 5} - {\it Threshold Eigenvalues}
\medskip

Let $H_0=-\Delta +V(x)$ in $L^2(\R^n),\  n>4$.  Assume $V(x)$ is
smooth and rapidly decaying for simplicity.  Then, under certain
conditions on the spectrum of $H_0$, and the behavior of its resolvent 
 at zero energy, one can prove local decay and $L^\infty$ decay with a
rate $r>2$; see \cite{kn:JK}, \cite{kn:JSS}.

In such cases,  it follows by Theorem 2.1  
 that a threshold eigenvalue at $\lambda =0$, if it exists, 
is unstable with respect to small  and generic 
perturbations, $W$.
\medskip

\noindent{\bf Example 6} -  {\it Stark Effect - atom in a uniform
electric field}

The Stark Hamiltonian is given by 
\be 
H\ =\ -\Delta + V(x) + \vec E\cdot x 
\nn
\ee
acting on $L^2(\R^n)$. If $V(x)$ is real valued  and not too singular,
then for $\vec E\ne0$ 
the continuous spectrum of $H$ is $(-\infty , \infty)$. To see this
apply Theorem 9.1 with
 $A\equiv \vec E\cdot p,\quad p=-i\nabla$. Thus, 
 if $H$ has an eigenvalue, it is necessarily  embedded in the 
 continuous spectrum. 

Our results can be used to show that any embedded eigenvalue is generically unstable
 ( {\it
i.e.} provided the Fermi golden rule resonance condition {\bf (W3)} holds) and perturbs to a  
resonance.

To see this, one can proceed by a
is decoupling argument; see \cite{kn:CoH}. This reduces the problem to a
direct sum of Hamiltonians, as in example 2, with Hamiltonians  of the
form:
\ba 
H_1\ &=&\ -\Delta\ +\ \vec E\cdot x\ +\ {\tilde V}(x,-i\nabla)\nn\\
H_2\ &=&\ -\Delta\ +\ V_b(x),\nn\\
W\  &=&\ {\tilde W}(x,-i\nabla)\nn
\ea
The strategy is then to use the techniques of appendix B to verify
hypotheses {\bf (W)} and the techniques of appendix D to prove the necessary 
 local decay esimates in {\bf (H)} for the  
the operator $H_0\ =\ \ {\rm diag}(H_1,H_2)$.
\medskip

\noindent{\bf Example 7} - {\it The Radiation Problem}
\medskip

The radiation problem is the fundamental problem which motivated work
on quantum resonances. See the work of Weisskopf \& Wigner \cite{kn:WW}, 
 following
 Dirac \cite{kn:Dirac} and Landau \cite{kn:Landau}. We present here a
very brief description of the problem and the relation to our methods.
For a more detailed discussion of the formulation see \cite{kn:BFS}.
 
The free Hamiltonian, $H_0$, is the direct sum operator acting on ${\cal
H}_a\oplus{\cal H}_{photon}$. Here, ${\cal H}_a$ is the Hilbert space
associated with an atom or molecule. ${\cal H}_{photon}$ is the Fock space
of free photons. $H_0$ is then the Hamiltonian of a decoupled particle
and free photon system:
\be
H_0\ =\ H_a\otimes I\ +\ I\otimes H_{photon}.\nn
\ee

The next step is to introduce the 
 interaction term $W$ that couples the photon-radiation field
to the atom. In quantum electrodynamics, this coupling is given 
 by the standard minimal coupling, but in  
general it is sufficient to consider a simple approximation
{\it e.g.} the dipole approximation \cite{kn:AE}. The goal is to show that all
eigenvalues of the original atom, except the ground state, are destabilized
by the coupling and become resonances. 
 This is the phenomenon of spontaneous emission.
 One is also interested in the computation of the {\it lifetime}
and the transition probabilities. 

A simplified Hamiltonian which incorporates the essential mathematical
features of the radiation problem is: 
\be
H\ =\ H_a\otimes I\ +\ I\otimes H_{photon}\ +\ \lambda W,
\nn
\ee
where $H_a\ =\ -\Delta\ +\ V(x)$ acting on $L^2(\R^n)$ describes the
atom, and the coupling is given by:
\be
W\ =\ \int\ \left(\omega(\vec k)\right)^{-1/2}
 \left( g(\vec k)\ e^{i\vec k x}
\ a_{\vec k}\ +\ {\overline g}(\vec k)\ e^{-i\vec
k\cdot\vec x}\ a_{\vec k}^\dagger\ \right)\ d\vec k.
\nn
\ee
The Hamiltonian associated with the photon field is given by:
\be
H_{photon}\ =\ \int\ \omega(\vec k)a^\dagger_{\vec k}a_{\vec k}\ 
 d\vec k
\nn
\ee
the second quantization of multiplication in Fourier space by
$\omega(\vec k)$ in $L^2(\R^n)$. 
Hence, $H_{photon}$ acts on the Fock space of bosons:
\be
{\cal F}\ =\ \oplus_{m=1}^\infty\ \otimes_{sym}^m L^2(\R^3),
\nn
\ee
where $\otimes_{sym}^m$ denotes the m-fold symmetric tensor product of 
 $L^2(\R^3)$.
The operator $a_{\vec k}^\dagger$ is the creation operator on ${\cal F}$
and $a_{\vec k}$, its adjoint. 

For realistic photons, we must have $g\equiv 1$ and 
   $\omega(\vec k)\sim |\vec k|$ for $\vec k$ near zero. However, to
make mathematical sense of the above Hamiltonian we need to introduce
the {\it ultraviolet cutoff}; 
\be g=0\quad {\rm for}\quad |\vec k |>>1.
\nn\ee 
When the coupling constant $\lambda$ is zero, and so $H=H_0$, it is fairly
easy to verify our conditions {\bf (H)} for $H_0$, even in the massless
photon case: $\omega(\vec k) = |\vec k|$\  
\cite{kn:BFS},\cite{kn:BFSS}. The conditions {\bf (W)}
however fail when $\omega(k) = |\vec k|$ since in this case 
 the  interaction $\lambda W$ is not localized.
On the other hand, in the massive case ($\omega(\vec k) =
\sqrt{m^2+|\vec k|^2},\ m\ne0$), the interaction is localized for quite
general $g(\vec k)$; see \cite{kn:G}. In this case, our conditions {\bf
(W)} can be verified and therefore the results of Theorem 2.1 can be
applied. 
\bigskip
\section{ Appendix A - Proof of Local Decay Proposition 2.1}
\bigskip

Our aim is to prove  local decay estimates for 
 $e^{-iH_0t}(H_0-\Lambda -i0)^{-1}P_c^\#$ 
using the given local decay estimates for $e^{-iH_0t}P_c^\#$, 
where $\Lambda\in {\bf T^\#}$.
The proof is split into two parts: analysis near $\Lambda$ and analysis
away from $\Lambda$.

Let $\Delta$ be a small interval about $\Lambda$ and $g_\Delta$ denote a
smoothed out characteristic function of $\Delta$ and ${\overline g}_\Delta
= 1 - g_\Delta$.
We write
\ba
&& e^{-iH_0t}(H_0-\Lambda -i0)^{-1}P_c^\# \nn\\
&=& e^{-iH_0t}(H_0-\Lambda -i0)^{-1}P_c^\#\, \left( g_\Delta +
{\overline g}_\Delta \right) \nn\\
&\equiv& T^t_\Delta + S^t_\Delta.
\nn
\ea
We first estimate the operator $T^t_\Delta$.
Let $\varepsilon >0$ and set 
$$T^t_{\Delta,\varepsilon} =  e^{-i(H_0-\Lambda -i\varepsilon ) t}
  (H_0-\Lambda -i\varepsilon)^{-1}P_c^\# g_\Delta.$$

Then, by {\bf (H4)} 
$$T^t_{\Delta,\varepsilon} = 
i\int_t^\infty\, e^{-i(H_0-\Lambda
-i\varepsilon ) s}\, P_c^\#\, g_\Delta\, ds.$$
Let $\la x\ra^\sigma h\in L^2$. Then,

\ba
|| \la x\ra^{-\sigma} T^t_{\Delta,\varepsilon}\, h||_2
&\le& 
 \int_t^\infty\, ||\, \la x\ra^{-\sigma} e^{-i(H_0-\Lambda
-i\varepsilon ) s}\, P_c^\#\, g_\Delta\, h\, ||_2 ds\nn\\
&\le&  
\int_t^\infty\, e^{-\varepsilon s}
 ||\, \la x\ra^{-\sigma} e^{-iH_0s} P_c^\#\, g_\Delta\, h\, ||_2 \ ds
\nn\\
&\le&
\int_t^\infty\, e^{-\varepsilon s} \la s\ra^{-r}\, 
||\, \la x\ra^{\sigma}h\, ||_2 \, ds\nn\\
&\le& 
C\la t\ra^{1-r}\, ||\, \la x\ra^{\sigma}h\, ||_2. 
\nn
\ea
Therefore, taking $\varepsilon\downarrow 0$, we get:  
$$|| \la x\ra^{-\sigma} T^t_\Delta\, h||_2\, \le\, 
 C\la t\ra^{1-r}\, ||\, \la x\ra^{\sigma}h\, ||_2.$$

To  estimate $S^t_\Delta$, we exploit that the energy is localized 
 away from $\Lambda$,  and so 
 the resolvent
$(H_0-\Lambda)^{-1}$ is bounded. 
\ba
\la x\ra^{-\sigma}  S^t_\Delta \ \la x\ra^{-\sigma}  &=&\ 
 \la x\ra^{-\sigma}\, e^{-iH_0t}\, 
 (H_0-\Lambda -i0)^{-1}\, P_c^\#\ {\overline g}_\Delta\ \la
 x\ra^{-\sigma}
 \nn\\
&=&\ \la x\ra^{-\sigma}\ e^{-iH_0t}\ P_c^\# \la x\ra^{-\sigma} \ \cdot\ 
 \la x\ra^\sigma\ (H_0-\Lambda -i0)^{-1}\ {\overline g}_\Delta\ 
   \la x\ra^{-\sigma}
 \nn\\
\label{eq:A1}
\ea
For the operator norm we then have the bound:
\be
\|\ \la x\ra^{-\sigma}  S^t_\Delta \ \la x\ra^{-\sigma}\ \|\ \le\ 
\|\ \la x\ra^{-\sigma}\ e^{-iH_0t}\ P_c^\#\ \la x\ra^{-\sigma}\ \|\cdot 
\ \|\ \la x\ra^\sigma\ (H_0-\Lambda -i0)^{-1}\ {\overline g}_\Delta\
  \la x\ra^{-\sigma}\ \|.
\label{eq:A2}
\ee
We bound the first factor in (\ref{eq:A2}) using the assumed  local decay
estimate {\bf (H4)}. The second factor is controlled as follows.  Note that

\ba
&&\la x\ra^\sigma\ (H_0-\Lambda -i0)^{-1}\ {\overline g}_\Delta\ 
\la x\ra^{-\sigma}\ =\ \la x\ra^\sigma\ (H_0+c)^{-1}\ 
 {\overline g}_\Delta\
\la x\ra^{-\sigma}\ \nn\\
\qquad\qquad &+&\ (\Lambda + c)\ \la x\ra^\sigma\ (H_0+c)^{-1}\ \la x\ra^{-\sigma}\ 
 \cdot \la x\ra^\sigma\ 
 (H_0-\Lambda -i0)^{-1}\ {\overline g}_\Delta\ \la x\ra^{-\sigma}.
\ea
Taking operator norms and using hypothesis {\bf (H5)} and Theorem 11.2 of appendix E we
obtain the following bound on the second factor in (\ref{eq:A2}).
\ba
&& \left(\ 1-\left|\Lambda + c\right|\|\la x\ra^\sigma\ (H_0+c)^{-1}\ \la x\ra^{-\sigma}\|\
\right)\ \|\ \la x\ra^\sigma\ (H_0-\Lambda -i0)^{-1}\ {\overline g}_\Delta\
\la x\ra^{-\sigma}\ \|\ \le\nn\\
&&\ \ \ \|\ \la x\ra^\sigma\ (H_0+c)^{-1}\
 {\overline g}_\Delta\
 \la x\ra^{-\sigma}\ \|\ \le\ \left(\ 1+ \|\ \la x\ra^\sigma\ g_\Delta \la
 x\ra^{-\sigma}\ \|\right)\  \|\ \la x\ra^\sigma\ (H_0+c)^{-1}\la x\ra^{-\sigma}\ \|
 \nn
 \ea
This completes the proof.

\section{ Appendix B:\, Operator norm estimates involving  $g_\Delta(H)$  
}
\bigskip
\def\tgdh{ {\tilde g}_\Delta (H) }
In this section we prove Propositions 3.1 and 4.1. These
 propositions
require some operator calculus. 

Let $\hat h(\lambda )$ denote the Fourier transform of the function $g$,
with the normalization:
$$\hat h (\mu )\, =\, (2\pi )^{-1}\int e^{i\mu \lambda }h(\mu)\, d\mu.$$
\medskip

\centerline {\it Proof of Proposition 3.1:}

  Recall that $\lambda_0$ denotes an embedded eigenvalue of the
unperturbed operator, $H_0$, $g_\Delta$ is a smoothed out
characteristic function
of the interval $\Delta$, and $I$ is an open set which contains the
support of $P_1$ and is disjoint from $\Delta$.
\medskip

We need to show that 
\be
B\, =\, (I-g_\Delta (H) \gI)^{-1}
\label{eq:Beqn}
\ee
is bounded and we do this by showing that $||g_\Delta (H) \gI ||$ has small
 norm. We use techniques \cite{kn:SS}. 

Let $\tilde\Delta$ be an interval which contains and is slightly larger
than $\Delta$. Then
\ba
g_\Delta (H) \gI &=& 
g_\Delta (H)\, \left( I\ -\ g_{\tilde \Delta}(H_0) \right)\, \gI\nn\\
&=& g_\Delta (H)\, g_{\Delta '} (H_0)\, \gI \nn\\
&=& g_\Delta (H)\,\left( g_{\Delta '} (H_0) - g_{\Delta '} (H) \right)\,
 \gI,
\ea
where $\Delta$ and $\Delta '$ are disjoint.

We now obtain, an expression for the above
difference, which is easily estimated. 
Using the Fourier transform we have that 
\be
 g_{\Delta '}(H_0) - g_{\Delta '}(H)\, =\, \int\, 
 (e^{i\mu H_0}-e^{i\mu H}) {\hat g}_{\Delta '}(\mu )\, d\mu.
\label{eq:B.1}
\ee
Furthermore,
\ba
e^{i\mu H_0}-e^{i\mu H} &=& (\ I\ -\ e^{i\mu H}e^{-i\mu H_0}\ )\, e^{i\mu H_0} 
\nn\\
&=& -\int_0^\mu\, {d\over ds} e^{isH}e^{-isH_0}\, ds\, e^{i\mu H_0}\nn\\ 
&=& -\int_0^\mu\, e^{isH}\, i(H-H_0)\, e^{-isH_0}\, ds\, e^{i\mu H_0}\nn\\
&=& -i\int_0^\mu\, e^{isH}\, W e^{-isH_0}\, ds\, e^{i\mu H_0}.
\label{eq:B.2}
\ea
Substitution of (\ref{eq:B.2}) into (\ref{eq:B.1}) yields:
\be
g_{\Delta '}(H_0) - g_{\Delta '}(H)\, =\, -i\int\,
 {\hat g}_{\Delta '}(\mu)\ e^{i\mu H}\ d\mu \, 
\int_0^\mu\, e^{-isH}\, W e^{isH_0}\, ds\ .
\label{eq:B.3}
\ee
We now apply the operator $g_\Delta (H)$ to the expression in (\ref{eq:B.3})
 and estimate:
\ba
&& ||\, g_\Delta (H)\, \left( g_{\Delta '}(H_0) - g_{\Delta '}(H)\right) \, ||\,  \nn\\
&\le& \int\, |\hat g_{\Delta '}(\mu )|\, \int_0^\mu\, ||g_{\Delta}(H)W||\, ds\,
d\mu\, \le\, \int\, |\hat g_{\Delta '}(\mu )|\, |\mu |\, d\mu\, ||g_{\Delta}(H)W||\nn\\
&\le&\, C\, |\Delta |^{-1}\, ||g_{\Delta}(H)W||\, \le C\, 
|\Delta |^{-1}\, |||W|||.
\label{eq:B4}
\ea
Therefore,
$$||\, g_\Delta (H)\gI\, ||\, \le\, C\, |\Delta |^{-1}\, |||W|||.
\label{eq:B5}$$ 
and $\left(I-g_\Delta (H)\gI\right)^{-1}$ is bounded provided
 $|\Delta |^{-1}\, |||W|||<\theta$ is sufficiently small\ ; see {\bf
(W4)}. 
\bigskip

\centerline {\it Proof of Proposition 4.1}

We estimate the norm of the  operator 
\be
{\cal G}\ =\ \la x\ra^\sigma  W\tilde g_\Delta(H)\la x\ra^\sigma
\nn
\ee
in
terms of $||| W |||$, defined in {\bf (W2)}.

Recall that by (\ref{eq:tildeg})
\be \tilde g_\Delta(H)\ =\ g_\Delta(H)\ 
						  \left( I-g_\Delta(H) g_I(H_0)\right)^{-1}\
						  {\overline g}_I(H_0).
\label{eq:gt}
\ee
Using (\ref{eq:gt}) we express ${\cal G}$ as the product of operators:
\ba
&& \la x\ra^\sigma W\tilde g_\Delta(H)\la x\ra^\sigma\ =\ {\cal G}_1\cdot
{\cal G}_2\cdot  {\cal G}_3\nn\\
&\equiv&\ \la x\ra^\sigma W g_\Delta(H)\la x\ra^\sigma\cdot\ 
\la x\ra^{-\sigma}\left[ I-g_\Delta(H) g_I(H_0)\right]^{-1}\la
x\ra^\sigma\cdot\ \la x\ra^{-\sigma} {\overline g}_I(H_0) \la x\ra^\sigma.
\label{eq:Gfactored}
\ea

Therefore it suffices to obtain upper bounds for $||{\cal G}_j||,\
j=1,2,3$. We shall use some general operator calculus estimates of
appendix E, especially Theorem 11.1.
\medskip

\nit{\bf Bound on ${\cal G}_3$:} This follows from Theorem 11.1 of
appendix E, with $A=H_0$ and $\varphi = 
 \overline g_I$, a function which is smooth and
rapidly decaying at infinity.

\medskip

\nit{\bf Bound on ${\cal G}_2$}:

By our hypotheses and the proof of Proposition 3.1, 
$|| g_\Delta(H) g_I(H_0)||$ is small and
\be \left( I-g_\Delta(H) g_I(H_0)\right)^{-1}\ =\ \sum_{n=0}^\infty
 \left( g_\Delta(H) g_I(H_0) \right)^n \nn\ee
 converges in norm.
We need  to show this in the weighted norms. For this, we will show that
the norm of $g_\Delta(H) g_I(H_0)$ is small in the weighted norm, {\it
i.e.}
\be \la x\ra^\sigma g_\Delta(H) g_I(H_0) \la x\ra^{-\sigma}\ =\ 
 {\cal O}\left( |||W|||\right)\nn\ee

Since the supports of  $g_\Delta$ and $g_I$ are disjoint 
\ba
\la x\ra^\sigma g_\Delta(H) g_I(H_0) \la x\ra^{-\sigma}\ &=&\ 
 \la x\ra^\sigma \left( g_\Delta(H) - g_\Delta(H_0) \right)
 g_I(H_0) \la x\ra^{-\sigma}\nn\\
 &=&\ \la x\ra^\sigma \left( g_\Delta(H) - g_\Delta(H_0) \right)\la
 x\ra^{-\sigma}\ \cdot\ \la x\ra^\sigma 
  g_I(H_0) \la x\ra^{-\sigma}.
  \nn\ea
  By parts (a) and (b), respectively, of Theorem 11.1  both 
  \ba
  &&\| \la x\ra^\sigma 
	g_I(H_0) \la x\ra^{-\sigma}\| < \infty\ {\rm and}\  \nn\\
 &&
\la x\ra^\sigma \left( g_\Delta(H) - g_\Delta(H_0) \right)\la
	 x\ra^{-\sigma}\ =\ {\cal O}\left( |||W|||\right).\label{eq:zzz}
	 \ea
 \medskip

 \nit{\bf Bound on ${\cal G}_1$:}
Expanding about the unperturbed operator, $H_0$, we have:
\ba
{\cal G}_1\ &=&\ 
 \la x\ra^\sigma W g_\Delta(H)\la x\ra^{\sigma}\ \nn\\ 
&=&\la x\ra^\sigma W (H+c)^{-1} \la x\ra^\sigma\ \cdot\ 
 \la x\ra^{-\sigma} (H+c)g_{\Delta}(H_0)\la x\ra^\sigma
\ea
Taking norms, we get 
\be
\| {\cal G}_1\|\ \le\   
  \|\la x\ra^\sigma W (H+c)^{-1} \la x\ra^\sigma\|\ \cdot\  
   \|  \la x\ra^{-\sigma} (H+c)g_{\Delta}(H_0)\la x\ra^\sigma\| 
  \label{eq:thing1}\ee
  Consider the first factor in (\ref{eq:thing1}).  We show that it is of
  order $ |||W||| $ as $|||W|||\to0$. 
  Note that 
  \ba
  \la x\ra^\sigma W (H+c)^{-1} \la x\ra^\sigma\ &=&\ \la x\ra^\sigma W
  (H_0+c)^{-1} \la x\ra^\sigma\nn\\
  \ &-&\ \la x\ra^\sigma W (H_0+c)^{-1} \la x\ra^{-\sigma}\ \cdot\ 
	\la x\ra^\sigma W (H+c)^{-1} \la x\ra^\sigma.\nn
	\ea
 Taking norms we obtain
 \be
 \left(\ 
 1 -\ \|\la x\ra^\sigma W (H_0+c)^{-1} \la x\ra^{-\sigma}\|
 \ \right)\ \| \la x\ra^\sigma W (H+c)^{-1} \la x\ra^\sigma\|\le
 \ \| \la x\ra^\sigma W (H_0+c)^{-1} \la x\ra^\sigma\|.
 \ee
Therefore, if $|||W||| < 1/2$ the first factor of  (\ref{eq:thing1}) is
bounded by $2|||W|||$. The second factor of  (\ref{eq:thing1} is bounded
by Theorem 11.1.
\medskip

   Finally, we note that 
	the above bounds on ${\cal G}_j$ complete the proof of Proposition
   4.1.

\def\adj{\mathop{\rm adj}}
\def\diag{\mathop{\rm diag}}
\def\rank{\mathop{\rm rank}}
\def\span{\mathop{\rm span}}
\def\rdots{\mathinner{\mkern1mu\raise1pt\vbox{\kern1pt\hbox{.}}\mkern2mu
   \raise4pt\hbox{.}\mkern2mu\raise7pt\hbox{.}\mkern1mu}}
\newcommand{\T}{{\rm T\kern-.35em T}}
\newcommand{\goth}[1]{#1}
\section{ Appendix C:\,  Expansion of the complex frequency, $\omega_*$
}

\bigskip
\def\tgdh{ {\tilde g}_\Delta (H) }
In this section we prove Proposition 3.3, in which an expansion of
the complex frequency, $\omega_*$, is presented. In particular, our
goal will be to obtain an expansion of $\omega_*$ which is explicit to
second order in the perturbation, $W$,  with an error term of order
$|||W|||^3$.

Recall that 
\be
   \omega_* = \lambda_0 + \omega_A - \omega_B,  
\label{eq:omegastarAB}
\ee
where
\be
\omega_A = \left( \psi_0, \; W\tilde{g}_\Delta(H)\psi_0 \right)\ =\
\omega_1-\lambda_0,\ ({\rm see\ } (\ref{eq:omega1})), 
\nn
\ee
and 
\be
\omega_B = \left( W\psi_0, \tilde{g}_\Delta(H) R_{H_0}(\omega + i0)
P^\#_c \;W
\,  \tilde{g}_\Delta(H) \psi_0 \right).
\nn
\ee

\noindent
 {\it Expansion of}  $\omega_A$:

\ba
\omega_A &\equiv& 
\left( \psi_0, \; W\tilde{g}_\Delta(H)\psi_0 \right) \nonumber \\
&=&  (\psi_0, W\psi_0) + \left( W \psi_0, \; B [g_\Delta(H) - g_\Delta(H_0)] \psi_0 \right)\nn \\
&\equiv& (\psi_0, \; W\psi_0) + \beta .
\label{eq:5}
\ea
In what follows, we shall frequently use the notation $(H-\lambda)^{-1}$
and ${1\over{H-\lambda}}$ interchangeably.

\noindent
\begin{prop}
\be
[g_\Delta(H) - g_\Delta(H_0)]\psi_0 = - \bar{g}_\Delta (H - \lambda_0)^{-1} W \psi_0
\label{eq:5a}
\ee
\end{prop}

\noindent
{\it proof:}  Noting that $H - H_0 = W$, we have the expansion formula
\begin{eqnarray*}
g_\Delta(H) &-& g_\Delta(H_0) = \int \hat{g}_\Delta (\lambda)\ \big( e^{i\lambda H} - e^{i\lambda H_0})\  d\lambda \\
&=& \int \hat{g}_\Delta(\lambda)\ e^{i\lambda H}\  (1 - e^{-i\lambda H} e^{i\lambda H_0})\ d\lambda \\
&=& i \int \hat{g}_\Delta(\lambda)\  e^{i\lambda H} \; \int^\lambda_0 e^{-isH} We^{isH_0}\ ds \; d\lambda .
\end{eqnarray*}

We next apply this expansion to $\psi_0$, where $H_0\psi_0 = \lambda_0 \psi_0$ and obtain
\ba
(g_\Delta(H) - g_\Delta(H_0))\psi_0 &=& 
 i \int \hat{g}_\Delta(\lambda)\ e^{i\lambda H}  \int^\lambda_0 e^{-isH}\ W\
  e^{is\lambda_0} \psi_0\  ds \; d\lambda \nn\\
&=& i \int \hat{g}_\Delta (\lambda) e^{i\lambda H} \int^\lambda_0 e^{-isH + is\lambda_0}\  W\ \psi_0\  ds d\lambda\nn \\
&=& i \int \hat{g}_\Delta(\lambda)\ e^{i\lambda H}\  {e^{-i\lambda H+i\lambda\lambda_0} - 1 \over {-iH + i\lambda_0}} W \psi_0 \ d\lambda \nn\\
&=& -\int \hat{g}_\Delta {e^{i\lambda\lambda_0} \over H-\lambda_0 }\  d\lambda\ W\ \psi_0 + 
 \int \hat{g}_\Delta(\lambda)   {e^{i\lambda H} \over H - \lambda_0}\ W\
 \psi_0\  d\lambda \nn\\
&=& -g_\Delta(\lambda_0) {1 \over H-\lambda_0} W\psi_0 + g_\Delta(H) {1 \over H-\lambda_0} W\psi_0 \nn\\
&=& -\big( 1 - g_\Delta(H) \big) {1 \over H-\lambda_0} W\psi_0,\quad
(g_\Delta(\lambda_0)=1)\nn\\
&\equiv& -\bar{g}_\Delta (H) (H-\lambda_0)^{-1} W\psi_0 .
\nn
\ea
 This completes the proof of the Proposition.  

Substitution of (\ref{eq:5a}) into the above expression for $\beta$ 
 yields
\be
\beta = - \left( W\psi_0, \; B\bar{g}_\Delta(H)(H - \lambda_0)^{-1}\; W\psi_0 \right) .
\label{eq:beta}
\ee
Let $h(\lambda)$ be a function which is equal to one on the support 
 of $\bar{g}_\Delta$
   and is zero outside  a small  neighborhood  of the support of
   $\bar{g}_\Delta$.
  Therefore $(H_0 - \lambda_0)^{-1}\, h(H_0)$ is bounded.
 A computation yields:

\begin{prop}
\ba
\omega_A &=& \left(\psi_0, W\psi_0\right) + \beta \nonumber \\
&=& \left(\psi_0, W\psi_0\right) - 
 \left(W\psi_0, \bar{g}_\Delta(H_0)(H_0 - \lambda_0)^{-1} \; W\psi_0\right) 
  \label{eq:5x} \\
 &-& \left(W\psi_0, \bar{g}_\Delta(H)(H-\lambda_0)^{-1} \;
  (h(H) - I) W\psi_0\right) 
  \nonumber \\
 &-& \left(W\psi_0, \bar{g}_\Delta(H_0)(H_0-\lambda_0)^{-1}
  [h(H) - h(H_0)]W\psi_0\right) 
  \nn \\
 &+& \left(W\psi_0, [g_\Delta(H)-g_\Delta(H_0)] h(H)
  (H-\lambda_0)^{-1}W\psi_0\right)\nn \\
 &+& \left(W\psi_0, \bar{g}_\Delta(H_0)(H_0-\lambda_0)^{-1} W(H-\lambda_0)^{-1} 
  h(H)W\psi_0\right) \nn\\
 &-& \left(W\psi_0, Bg_\Delta(H) g_I(H_0) \bar{g}_\Delta(H)
 (H-\lambda_0)^{-1} W\psi_0\right).
\ea
\end{prop}

Note also that the second term in (\ref{eq:5x}) can be expressed as
\ba
&&( W\psi_0, \bar{g}_\Delta(H_0)(H_0 - \lambda_0)^{-1} \; W\psi_0)\ =\nn\\
&& (W\psi_0, \bar{g}_\Delta(H_0)(H_0 - \omega)^{-1} \; W\psi_0)\ 
-\  (W\psi_0, \psi_0) 
\cdot (W\psi_0, \bar{g}_\Delta(H_0)(H_0 - \lambda_0)^{-1}
 (H_0 - \omega)^{-1} W\psi_0 )\nn\\
\label{eq:6}
\ea

\noindent
{\it Expansion of} $\omega_B$: 

\ Let $R_{H}(\lambda)\ \equiv\ (H-\lambda)^{-1}$.
Recall that $\omega_B$ is given by the expression:
$$\omega_B = \left( W\psi_0, \tilde{g}_\Delta(H) R_{H_0}(\omega + i0) P^\#_c \;W\,  \tilde{g}_\Delta(H) \psi_0 \right),
$$
and  
$\tilde{g}_\Delta(H) = Bg_\Delta(H)(I - P_1)$. 
 We find after some computation:
\begin{prop}
\ba
\omega_B &=& (W\psi_0, g_\Delta(H_0) R_{H_0}(\omega + i0) P^\#_c \; W\psi_0) 
 \nn \\
&+& (W\psi_0, [B-I] g_\Delta(H_0) R_{H_0}(\omega + i0)P^\#_c \; 
                 WBg_\Delta(H) \psi_0)
\nn \\
&+& (W\psi_0, [g_\Delta(H)-g_\Delta(H_0)]R_{H_0}(\omega + i0) P^\#_cW
{\tilde g}_\Delta(H)\psi_0)\nn\\
&+& (W\psi_0, g_\Delta(H_0) R_{H_0}(\omega + i0) P^\#_c WB
		 [g_\Delta(H)-g_\Delta(H_0)]\psi_0) \nn \\
&+& (W\psi_0, (B-I)[g_\Delta(H) - g_\Delta(H_0)]{\overline g}_I(H_0) 
 R_{H_0}(\omega + i0) P^\#_c WBg_\Delta(H) \psi_0). 
\label{eq:7}
\ea
\end{prop}
Here, we have used that $B\psi_0=\psi_0$.
 More generally,  
 $(B-I)g_\Delta(H_0)=Bg_\Delta(H)g_I(H_0)g_\Delta(H_0)=0,$ and therefore
the second term in (\ref{eq:7}) is zero.

It follows from (\ref{eq:5}), (\ref{eq:6}) and (\ref{eq:7}) that
\ba
\omega_* &=& \lambda_0 + \omega_A - \omega_B \nn\\
&\equiv& \lambda_0 + (\psi_0, W\psi_0)\ -\ \left(\Lambda +
i\Gamma\right)\ +\ \sum^9_{j=1} E_j,\label{eq:8}
\ea
where
\be
\Lambda + i\Gamma\ =\ 
(W\psi_0, \bar{g}_\Delta(H_0)(H_0 - \omega)^{-1} W\psi_0) + 
 (W\psi_0, g_\Delta(H_0)
  (H_0-\omega - i0)^{-1} P^\#_c W\psi_0).\label{eq:LamGam} 
 \ee
and
\ba
E_1 &=& (W\psi_0,\psi_0) \cdot (W\psi_0, \bar{g}_\Delta
 (H_0 - \lambda_0)^{-1}(H_0 - \omega)^{-1} W\psi_0)\nn\\
E_2 &=& (W\psi_0, (H -\lambda_0)^{-1} (h(H)-I) \bar{g}_\Delta(H) W\psi_0)\nn\\
E_3 &=& - (W\psi_0, \bar{g}_\Delta(H_0)(H_0 -\lambda_0)^{-1}[h(H) - h(H_0)]
 W\psi_0)\nn\\
E_4 &=&  (W\psi_0, [g_\Delta(H) - g_\Delta(H_0)] h(H) (H -\lambda_0)^{-1}
 W\psi_0)\nn\\
E_5 &=&  (W\psi_0, \bar{g}_\Delta(H_0) (H_0 -\lambda_0)^{-1} W
 (H -\lambda_0)^{-1}
      h(H) W\psi_0)\nn\\
E_6 &=& (W\psi_0, Bg_\Delta(H)g_I(H_0)
 [g_\Delta(H_0) - g_\Delta(H)]\bar{g}_I(H_0) R_{H_0} (\omega + i0) P^{\#}_c 
  WBg_\Delta(H)\psi_0)\nn\\
&&\nn\\
E_7 &=& (W\psi_0, g_\Delta(H_0) R_{H_0}(\omega + i0) P^{\#}_c WB 
 [g_\Delta(H_0) - g_\Delta(H)]\psi_0)\nn\\
E_8 &=& (W\psi_0, [g_\Delta(H_0) - g_\Delta(H)] R_{H_0} (\omega + i0) 
 P^{\#}_c WBg_\Delta(H)\psi_0)\nn\\
E_9 &=& (W\psi_0, B[g_\Delta(H_0) - g_\Delta(H)]
         g_I(H_0)\bar{g}_\Delta(H) (H-\lambda_0)^{-1} W\psi_0)
		 \nn
\ea

We now claim that the terms $E_j,\quad j=1,...,9$ are all of
order $||| W|||^3$. Consider first $E_1=E_1^a\cdot E_1^b$.
Estimation of the first factor gives:
\be
|E_1^a|\  
\le C\ |||W|||,
\nn
\ee
by Proposition 4.1.

Estimation of the second factor gives:
\ba
|E^b_1|\ &=&\  \left|\left( Wg_\Delta(H_0)\psi_0,{\overline
g}_\Delta(H_0)(H_0-\lambda_0)^{-1}(H_0-\omega)^{-1} 
 Wg_\Delta(H_0)\psi_0\right)\right|
\nn\\
&\le&\ \|\la x\ra^\sigma Wg_\Delta(H_0)\psi_0\|_2^2\ 
 \|\la x\ra^{-\sigma}{\overline g}_\Delta(H_0)(H_0-\lambda_0)^{-1} 
 (H_0-\omega)^{-1}\la x\ra^{-\sigma}\|\nn\\
 &\le&\ C\ |||W|||^2\ \|\la x\ra^{-\sigma}{\overline
 g}_\Delta(H_0)(H_0-\lambda_0)^{-1}
  (H_0-\omega)^{-1}\la x\ra^{-\sigma}\| \le\ C\ |||W|||^2,
  \nn\ea
  by Theorem 11.1. Therefore, $|E_1|\le C\ |||W|||^3$.

The term $E_2$ is zero; $(h-1)\bar{g}_\Delta\equiv 0$ since $h\equiv 1$
on the support of $\bar{g}_\Delta$.

The term $E_5$ can be treated by the same type of estimates as $E_1$.
The remaining terms are $E_j,\ j=2,3,4,6,7,8,9$.  Each of these
expressions has two explicit occurrances of the perturbation, $W$, as
well as a difference of operators: $g_\Delta(H)-g_\Delta(H_0)$ or 
 $h(H)-h(H_0)$. By (\ref{eq:zzz}), these differences 
 are ${\cal O}(|||W|||)$, so we expect each of these terms to be 
 ${\cal O}(|||W|||^3)$. We carry this argument out for the term $E_7$.
 The other terms are similarly estimated. 

Consider $E_7$. Let $\tilde\Delta$ be an interval properly containing
$\Delta$ so restricted to the interval 
 $\Delta$,  $g_{\tilde\Delta}\equiv 1$ and 
 $g_\Delta = g_\Delta g_{\tilde\Delta}$. Then,
\ba
&& |E_7| = \left|\left(\la x\ra^\sigma g_\Delta(H)W\psi_0,
\ \la x\ra^{-\sigma}g_{\tilde\Delta}(H)R_0(\omega +i0)P_c^\#\la
x\ra^{-\sigma}\ \cdot \la x\ra^\sigma W B
[g_\Delta(H_0)-g_\Delta(H)]\psi_0\right)\right|
\nn\\
&\le&\ \left|\left(\la x\ra^\sigma g_\Delta(H)W\psi_0,
\ \la x\ra^{-\sigma}g_{\tilde\Delta}(H_0)R_0(\omega +i0)P_c^\#\la
x\ra^{-\sigma}\ \cdot \la x\ra^\sigma W B
[g_\Delta(H_0)-g_\Delta(H)]\psi_0\right)\right|
\nn\\
&& + \left|\left(\la x\ra^\sigma g_\Delta(H)W\psi_0,
\ \la x\ra^{-\sigma}\left[ g_{\tilde\Delta}(H)-g_{\tilde\Delta}(H_0)\right]
 \la x\ra^\sigma\ \la x\ra^{-\sigma}  
 R_0(\omega +i0)P_c^\#\la 
x\ra^{-\sigma}\ \cdot\right.\right.\nn\\
&& \ \ \ \ \ \ \ \ \ \  \left.\left.\ \la x\ra^\sigma W B
[g_\Delta(H_0)-g_\Delta(H)]\psi_0\right)\right|
\ea
 Using Proposition 4.1, Theorem 11.1 and (\ref{eq:zzz}) we have that
 \be
 |E_7|\ \le\ C\ |||W|||^3\ \|\la x\ra^{-\sigma}(H_0 - \omega - i0)^{-1}
  P_c^\# \la x\ra^{-\sigma}\|.
 \nn\ee
 That the term $ \|\la x\ra^{-\sigma}(H_0 - \omega - i0)^{-1}\la
  x\ra^{-\sigma}\|$ is finite is a consequence of Proposition 2.1 with
  $t=0$.
 Thus we have the
following proposition from which Proposition 3.4
follows.

\begin{prop}
\ba
&&(1)\  \Lambda + i\Gamma = 
   (W\psi_0, {\rm P.V.} (H_0 - \omega)^{-1} W\psi_0)\ +\  
 i\pi(W\psi_0, \delta(H_0 - \omega)(I - P_0) W\psi_0)\nn\\
&&(2)\  |E_j| \le C |||W|||^3 \ \ \ \ 1 \le j \le 9\nn\\
&&(3)\  \omega_* = \lambda_0 + (\psi_0, W\psi_0) - \Lambda - i\Gamma + E(W),\  
  {\rm where}\   |E(W)| \le C|||W|||^3.
 \ea
\end{prop}
It remains to verify part (1). 
This follows from an application of the 
 well-known distributional identity 
\be
(x\mp i0)^{-1}\ \equiv\ \lim_{\varepsilon\to0^+}(x\mp i\varepsilon)^{-1}\ =\ 
 {\rm P.V.}\ x^{-1}\ \pm\ i\pi\delta(x)
\nn
\ee
to the second term in equation (\ref{eq:LamGam}) and the identity   
 $g_\Delta(H_0)P_c^\#= I-P_0$. 
\section{Appendix D:\ General approach to local decay estimates}

\bigskip
Hypothesis {\bf (H4)} for our main theorem is one requiring that our
unperturbed operator, $H_0$, satisfy a suitable local decay estimate,
 (\ref{eq:ld1}). In this section we give an outline to a
very general approach to obtaining such estimates based on a technique
originating in the work of Mourre \cite{kn:Mr}; see also \cite{kn:PSS}.
 In the following general
discussion  
we shall  let   
 $H$ denote self-adjoint operator on a Hilbert space, ${\cal H}$,
keeping in mind that our application is to the unperturbed operator $H_0$.
 Let $E\in\sigma(H)$, and assume that an operator $A$ can be found such
that $A$ is self-adjoint and ${\cal D}(A)\cap {\cal H}$ is dense in
${\cal H}$.
Let $\Delta$ denote an open interval with compact closure.
We shall use the notation:
\be
{\rm ad}_A^n(H)\ =\ [\cdots[H,A],A],\cdots A],\nn
\ee
for the n-fold commutator.

 Assume the two conditions:  
  
  \nit {\bf (M1)}\  The operators  
\be g_\Delta(H)\ {\rm ad}_A^n(H)\ g_\Delta(H),\ \ 1\le n\le N\ee
can all be extended to a bounded operator on ${\cal H}$. 

\bigskip

\nit {\bf (M2)}\ {\it Mourre estimate}:
\be
g_\Delta (H)\ i[H,A]\ g_\Delta (H)\ \ge \theta\ g_\Delta(H)^2\ +\ K
\ee
for some $\theta >0$ and compact operator, $K$.
\bigskip
\bigskip

\begin{theo} (Mourre;\  see \cite{kn:CFKS} Theorem 4.9)

\nit Assume conditions {\bf (M1)-(M2)}, with $N=2$. Then, in the interval
$\Delta$, $H$ can only have 
absolutely continuous spectrum with finitely many eigenvalues
of finite multiplicity. Moreover, the operator
\be 
\la A\ra^{-1}\ g_\Delta(H)\ (H-z)^{-1}\ \la A\ra^{-1} 
\ee
is uniformly bounded in $z$, as an operator on ${\cal H}$.
If $K=0$, then there are no eigenvalues in the interval $\Delta$.
\end{theo}

\bigskip
\bigskip

\begin{theo} (Sigal-Soffer;\ see \cite{kn:SS},\cite{kn:GS},
\cite{kn:HS})

\nit Assume conditions {\bf (M1)-(M2)} with  $N\ge2$ and 
 $K=0$. Then, for all $\varepsilon >0$
\be
||\ F\left({|A|\over t}<\theta\right)\ e^{-iHt}\ g_\Delta (H)\psi\ ||_2
\ \le\ C \la t\ra^{-{N\over2}+\varepsilon}\ \|\ |A|^{N\over2}\psi\ \|_2,
\ee
and therefore
\be
||\ \la A\ra^{-\sigma}\ e^{-iHt}\ g_\Delta (H)\psi\ ||_2\ 
  \le\ C\ \langle t\rangle^{-\sigma}\ \|\ |A|^{N\over2}\psi\ \|_2,
\ee
for $\sigma < N/2$. Here, $F$ is a smoothed out characteristic
function, and  $F\left({|A|\over t}<\theta\right)$ is defined by
the spectral theorem.
\end{theo}

 Let $\Delta_1$ denote an open interval
 containing the closure  of $\Delta$.

\begin{cor}
 Assume that $\la x\ra^{-\sigma}\ g_{\Delta_1}(H)\la
 A\ra^\sigma$ is  
bounded. Then, in the above theorems we can replace the weight 
 $\la A\ra^{-\sigma}$ by $\la x\ra^{-\sigma}$.
\end{cor}
\bigskip

The strategy for using the above results to prove local decay estimates like
that in {\bf (H4)} is as follows. 
 Then 
\ba
\|\ \la x\ra^{-\sigma}\ e^{-iHt}g_\Delta(H)\psi\ \|_2
\ &=& \|\  \la x\ra^{-\sigma}g_{\Delta_1}(H) e^{-iHt}g_\Delta(H)\psi\|_2
\nn\\
\ &=& \|\  \la x\ra^{-\sigma}g_{\Delta_1}(H) \la A\ra^\sigma\cdot  
\la A\ra^{-\sigma}
\ e^{-iHt}g_\Delta(H)\psi\ \|\nn\\
&\le&\ \|\  \la x\ra^{-\sigma}g_{\Delta_1}(H) \la A\ra^\sigma \| \cdot
 \|\ \la A\ra^{-\sigma}
 \ e^{-iHt}g_\Delta(H)\psi\ \|\nn\\
 &\le&\ C\ \|\ \la A\ra^{-\sigma}
  \ e^{-iHt}g_\Delta(H)\psi\ \|\nn\\
 &\le&\ C_1 \|\ F\left({|A|\over t} <\theta\right)\ \la A\ra^{-\sigma}
  \ e^{-iHt}g_\Delta(H)\psi\ \|_2\ \nn\\
  &&\  +\ C_2\|\ F\left({|A|\over t}\ge \theta\ \right) \la A\ra^{-\sigma}
   \ e^{-iHt}g_\Delta(H)\psi\ \|_2
\label{eq:split}
\ea
Theorem 10.2 is used to obtain the decay of the first term on the right
hand side of (\ref{eq:split}), while we can replace $|A|$ by $\theta t$
in the second term.

\bigskip
\nit{\bf Remark:} Here we return to our comment in the introduction
on  the relation between our assumption {\bf (H4)} 
  (local decay for $e^{-iH_0t}$) and the hypothesis of dilation
analyticity, used in previous works. Dilation analyticity or its generalization,
 analytic deformation, is the requirement that the map:
\be 
d(\theta):\ \theta\mapsto\ \left(\ e^{i\theta A}\ H_0\ e^{-i\theta A}f,\
f\ \right),\ 
\ee
has analytic continuation to a strip, for $f$ in a dense subset of
${\cal H}$. Since the $n^{th}$  derivative of
$d(\theta)$  at
$\theta = 0$  is $\left({\rm ad}_A^n(H_0)f,f\right)$, 
 by the above local decay result, the
assumption {\bf (H4)} is the requirement that the
mapping, $d(\theta)$ be of class $C^3$.

\section{\bf Appendix E: Weighted norm estimates for functions of operators}

In appendices A, B and C we frequently require facts and estimates of
functions of a self-adjoint operator. In this section we give some
basic definitions and provide the statements and proofs of such
estimates. We shall refer to certain known results and our basic
references are \cite{kn:RS4} and \cite{kn:AMG}.

Let $A$ denote a self-adjoint operator with domain ${\cal D}$ which is
dense in a Hilbert space ${\cal H}$. 
Then we have that for any bounded continuous complex-valued function,
 $\varphi\in L^1(\R)$:
\be
\varphi (A)\ =\ {\rm weak}-\lim_{\varepsilon\downarrow 0} \pi^{-1}\int\ \varphi(\lambda)\ 
				   \Im\ R_A(\lambda +i\varepsilon)^{-1}\ d\lambda,
\label{eq:phiofA}
\ee
where  
$R_A(\lambda)\ =\ (A-\lambda)^{-1}$
 denotes the resolvent of $A$. Here and throughout this section all regions of
integration are assumed to be over $\R$ unless explicitly stated
otherwise.

\begin{theo} Let $\tA$ and $\tB$ denote bounded self-adjoint operators,
and let $\Gamma$ be a contour in the complex plane, not passing
through the origin, surrounding $\sigma(\tA)\cup\sigma(\tB)$ and lying
in the strip $|\Im\zeta |<1$. 

\nit (a) Let $\psi :\R\to\C$ be a $W^{2,1}$ function. 
Suppose 
\be
\eta_{\tA}\ \equiv\ \|\la x\ra^\sigma \tA \la x\ra^{-\sigma}\|\ <\
 {1\over2}\min \left[ {\rm distance}(\Gamma,0),1 \right]\label{eq:tAsmall}
\ee
Then, there exists a positive number
 $C_1=C_1\left( \|\psi\|_{W^{2,1}},\eta_{\tA} \right)$
such that 
\be
\|\la x\ra^\sigma\ \psi(\tA)\ \la x \ra^{-\sigma}\|\ <\ C_1.
\label{eq:psithing}
\ee
\nit (b) 
Let $\psi$ be as in part (a). 
Assume that $\tA$ and $\tB$ both satisfy condition (\ref{eq:tAsmall}). 
Then, there is a constant 
 $C_2=C_2\left(\|\psi\|_{W^{2,1}},\eta_{\tA},\eta_{\tB}\right)$ 
 such that
\be 
\|\la x\ra^\sigma\left[ \psi(\tA) - \psi(\tB)\right]
 \la x\ra^{-\sigma}\|\ \le\ C_2\ \|\la x\ra^\sigma\ 
 ( \tA-\tB)\ \la x\ra^{-\sigma}\|.
\label{eq:differencething}
\ee
\end{theo}
\medskip

The following result shows that the case of unbounded self-adjoint
 operators is reducible to Theorem 11.1.

 \begin{theo}
Suppose that Theorem 11.1 holds, and  
let $\Gamma$ and $\varphi$ be as in Theorem 11.1.
Furthermore, assume that
 $x^2\varphi''(x)$ and $x\varphi'(x)$ are $L^1$ functions.
 Let $A$ and $B$ be  
 densely defined 
 self-adjoint operators for which $(A+c)^{-1}$ and $(B+c)^{-1}$ are
  bounded for some real
number $c$ and satisfy the estimate (\ref{eq:tAsmall}).
 Then, 
 \be
 \|\la x\ra^\sigma\ \varphi(A)\ \la x \ra^{-\sigma}\|\ <\ C_1(\psi).
 \label{eq:varphithing}
 \ee
 and 
\be
\|\la x\ra^\sigma\left[ \varphi(A) - \varphi(B)\right]
 \la x\ra^{-\sigma}\|\ \le\ C_2(\psi)\ \|\la x\ra^\sigma\
  ( A-B)\ \la x\ra^{-\sigma}\|,
  \label{eq:varphidifferencething}
  \ee
  where the constants $C_1(\psi)$ and $C_2(\psi)$ are as in Theorem
  11.1, with  $\psi(x)\equiv \varphi(x^{-1}-c)$.
 \end{theo}
\medskip

\nit{\it proof:}
Let $\tA=(A+c)^{-1}$ and note that  
$\varphi(A)\ =\  \varphi(\tA^{-1}-c)\ =\ \psi(\tA).$ 
It suffices to show that $\psi(x)=\varphi(x^{-1}-c)$ satisfies the
hypotheses of Theorem 11.1.  
 It is simple to check that
 $\D_x^j\psi(x)\in L^1$, for $j=0,1,2.$  This proves Theorem 11.2.

 We now embark on the proof of Theorem 11.1. 
 A key tool is an expansion formula for
 $\varphi(A)$; see Proposition 6.1.4 on page 239 of \cite{kn:AMG}.

\begin{theo} 
Let $A$ be a densely defined self-adjoint operator 
 and $\varphi$ be as in the statement of Theorem 11.1. Then,

\ba
\varphi (A)\ &=&\ {1\over\pi}\ \int\ \varphi(\lambda)\ \Im\ R_A(\lambda
                                                   + i)\ d\lambda \nn\\
             &+&\ {1\over\pi}\ \int\ \varphi'(\lambda)\	
						 \Im\ iR_A(\lambda + i)\ d\lambda \nn\\
             &+&\  {1\over\pi}\ \int_0^1\ \tau\ d\tau\ 
					   \int\ \varphi''(\lambda)\
						 \Im\ i^2R_A(\lambda + i\tau)\ d\lambda\nn\\
             &\equiv&\ \varphi_1\ +\ \varphi_2\ +  \varphi_3. 
\label{eq:phiexpand}
\ea
where all integrals exist in the norm of the space of bounded operators
on ${\cal  H}$.
\end{theo}
\medskip

To prove Theorem 11.1 we first obtain a simple expression for the third
summand in (\ref{eq:phiexpand}) by interchanging order of
integration. We begin with a calculation of the $\tau -$ integral:
\ba
\int_0^1\ \tau\ d\tau\ \Im\ \left(\ i^2\ R_A(\lambda +i\tau )\ \right) 
\ &=&\ -\int_0^1\ \tau\ {1\over{2i}}\left(\ R_A(\lambda + i\tau)\ -\ 
	R_A(\lambda - i\tau)\ \right)\nn\\
\ &=&\ -\int_0^1\ d\tau\ \tau^2\ \left[\ (A-\lambda)^2+\tau^2\ \right]^{-1}\nn\\
\ &=&\ f(A;\lambda)\ -\ 1,\nn
\ea
where
\be
f(z;\lambda )\ =\ (z-\lambda)\ \int_0^{1\over{z-\lambda}}\ (1+\mu^2)^{-1}\ d\mu.
\label{eq:f-def} 
\ee
For each $\lambda$ in the support of $\varphi$, the function $f(z;\lambda)$ is 
analytic in the strip $|\Im z|<1$; this corresponds to choice an
appropriate branch of the function
$z\mapsto (z-\lambda)\arctan{(z-\lambda)^{-1}}$.
 By  (\ref{eq:phiexpand}) 
\be
\varphi_3(A) =\ {1\over\pi}\ \int\ \varphi''(\lambda)
f(A;\lambda)\ d\lambda.
\label{eq:arctanterm}
\ee

The strategy is as follows. 
\medskip

First, we observe that   
 $\la x\ra^\sigma\ \varphi_j(\tA)\ \la x\ra^{-\sigma}$ is bounded for
 $j=1,2$. This is true because $\varphi, \varphi'\in L^1$ and
 (\ref{eq:tAsmall}) can be used to bound the weighted norm of the 
  resolvent by 
 a convergent geometric series. 
 Therefore, it remains to bound the operator $\varphi_3(\tA)$, where
 $\varphi_3$ is given
 explicitly (\ref{eq:arctanterm}).

 \begin{lem}
 Let $\tA$ and $\tilde B$ denote  bounded self-adjoint operators and
  $f(\zeta)$ be a function which is defined and analytic in a
 neighborhood of $\sigma(\tA)\cup\sigma(\tilde B)$. 
  Let $\Gamma$ be a smooth contour in the
 domain of analyticity of $f$, surrounding $\sigma(\tA)\cup\sigma(\tilde
 B)$, not
 passing through the origin and  
 such that the estimate:
 \be
\| \la x\ra^\sigma\ M\  \la x\ra^{-\sigma}\| \le\  
{1\over2}\min_{\zeta\in\Gamma}|\zeta|,
\label{eq:onehalfhypothesis}
\ee
holds with $M=\tA$ and $M=\tilde B$.
 Then, there exist positive constants $C_1$ and $C_2$ such that  
 \ba \|\la x\ra^\sigma f(\tilde A)  \la x\ra ^{-\sigma} \|
 \ &\le& C_1
  \label{eq:l1}\\
  \|\la x\ra^\sigma\left[ f(\tilde A) - f(\tilde B)\right]\ \la
  x\ra^{-\sigma}\|\ &\le&\  C_2\ \|\la x\ra^\sigma 
		(\tilde A-\tilde B) \la x\ra^{-\sigma}\|
\label{eq:l2}
 \ea
 \end{lem}

 \nit{\it proof:} By the Cauchy integral formula we have 
 \be
  f(\tilde A)\ =\ 
  (2\pi i)^{-1}\int_\Gamma\ f(\zeta)\ (\tilde A-\zeta I)^{-1}\ d \zeta.
  \label{eq:foftA}
\ee
Part (a) follows by use of  (\ref{eq:onehalfhypothesis}) to expand the
resolvent in a geometric series and by termwise estimation in the
weighted norm.

\nit Part (b) follows by the same method; by (\ref{eq:foftA}) 
 applied to $\tilde B$ and
computation of the difference, we get:
\ba
f(\tilde A)-f(\tilde B)\ &=&\
 (2\pi i)^{-1}\int_\Gamma\ f(\zeta)\left[ (\tilde A-\zeta I)^{-1}-
  (\tilde B-\zeta I)^{-1}\right]\ d\zeta\nn\\
 &=& (2\pi i)^{-1}\int_\Gamma\ f(\zeta)\left[ (\tilde A-\zeta I)^{-1}  
\ (\tilde A-\tilde B)\ (\tilde B-\zeta I)^{-1}\right]\ d\zeta.\nn
\ea
Estimation in the weighted space yields (\ref{eq:l2}).
This completes the proof of the Lemma.
\medskip

  To complete the proofs of Theorems 11.1 and Theorem 11.2, 
  we need to estimate the operator 
  $\la x\ra^\sigma \varphi_3(\tA)\la x\ra^{-\sigma}$, 
  where $\tA$ is the bounded self-adjoint operator defined by 
   $\tA=(A+c)^{-1}$.
  We accomplish this by applying the previous lemma to the function
  $f(\zeta;\lambda)$ defined in (\ref{eq:f-def}), 
  where $\lambda$ is in the support of $\varphi$. 
    The function $f(\zeta;\lambda)$ is  analytic in the strip  
   $|\Im\zeta |<1$, and $\Gamma$ is, by hypothesis, a contour in its  
   domain of analyticity, surrounding 
   $\sigma(\tA)$ (respectively, $\sigma(\tA)\cup\sigma(\tilde B)$,) 
	and so that 
   (\ref{eq:onehalfhypothesis}) holds. Then, by Lemma 11.1 we have
   that $f(\tilde A;\lambda)$ and $f(\tilde B;\lambda)$ satisfy 
   (\ref{eq:l1}) and (\ref{eq:l2}). Finally, using the representation 
 formula for $\varphi_3$, (\ref{eq:arctanterm}), we have 
   \ba
   \| \la x\ra^\sigma \varphi_3(\tA)\la x\ra^{-\sigma}\|\ &\le&\
	 C_1\ \|\varphi''\|_{L^1},\nn\\
   \| \la x\ra^\sigma
   \left[\varphi_3(\tA)-\varphi_3({\tilde B})\right]  
   \la x\ra^{-\sigma}\|\ &\le&\ C_2\ \|\varphi''\|_{L^1}\ 
	\| \la x\ra^\sigma \left[ \tA - \tilde B\right] \la x\ra^{-\sigma}\|. 
   \ea
   This completes the proof.

\enddocument
\end

\enddocument
\end